\documentclass{article}

\usepackage{microtype}
\usepackage{graphicx}
\usepackage{subcaption}
\usepackage{booktabs} 

\usepackage{hyperref}

\usepackage[numbers,sort&compress]{natbib}

\usepackage[preprint]{icml2026}

\usepackage{amsmath}
\usepackage{amssymb}
\usepackage{mathtools}
\usepackage{amsthm}
\usepackage{booktabs}
\usepackage{microtype}
\usepackage{siunitx}
\usepackage{pdfpages}
\usepackage{xcolor}
\usepackage{cuted}
\usepackage{dblfloatfix}
\usepackage{placeins}

\usepackage[capitalize,noabbrev]{cleveref}

\theoremstyle{plain}

\theoremstyle{definition}

\theoremstyle{remark}

\definecolor{final}{RGB}{232, 0, 11}
\definecolor{enigma}{RGB}{2, 62, 255}
\definecolor{atm}{RGB}{255, 124, 0}
\definecolor{enigma+atm}{RGB}{26, 201, 56}
\definecolor{perceptogram}{RGB}{139, 43, 226}

\usepackage[textsize=tiny]{todonotes}

\icmltitlerunning{\textbf{ENIGMA}: EEG-to-Image in 15 Minutes Using Less Than 1\% of the Parameters}

\begin{document}

\twocolumn[
  \icmltitle{\textbf{ENIGMA}: EEG-to-Image in 15 Minutes Using Less Than 1\% of the Parameters}



  \icmlsetsymbol{equal}{*}

  \begin{icmlauthorlist}
    \icmlauthor{Reese Kneeland}{aj}
    \icmlauthor{Wangshu Jiang}{aj,wl}
    \icmlauthor{Ugo Bruzadin Nunes}{aj}
    \icmlauthor{Paul S. Scotti}{sophont,princeton}
    \icmlauthor{Arnaud Delorme}{ucsd}
    \icmlauthor{Jonathan Xu}{aj}
    
  \end{icmlauthorlist}

\icmlaffiliation{aj}{Alljoined}
  \icmlaffiliation{wl}{University of Waterloo}
  \icmlaffiliation{sophont}{Sophont}
  \icmlaffiliation{princeton}{Princeton Neuroscience Institute}
  \icmlaffiliation{ucsd}{University of California San Diego}

  \icmlcorrespondingauthor{Reese Kneeland}{reese@alljoined.com}

  \icmlkeywords{EEG, brain decoding, brain-computer interface, image reconstruction, diffusion}

  \vskip 0.3in
]



\printAffiliationsAndNotice{}  


\begin{strip}\centering
\vspace{-60pt}       
\setlength{\abovecaptionskip}{0pt}    
\setlength{\belowcaptionskip}{-10pt}    
\includegraphics[width=\textwidth]{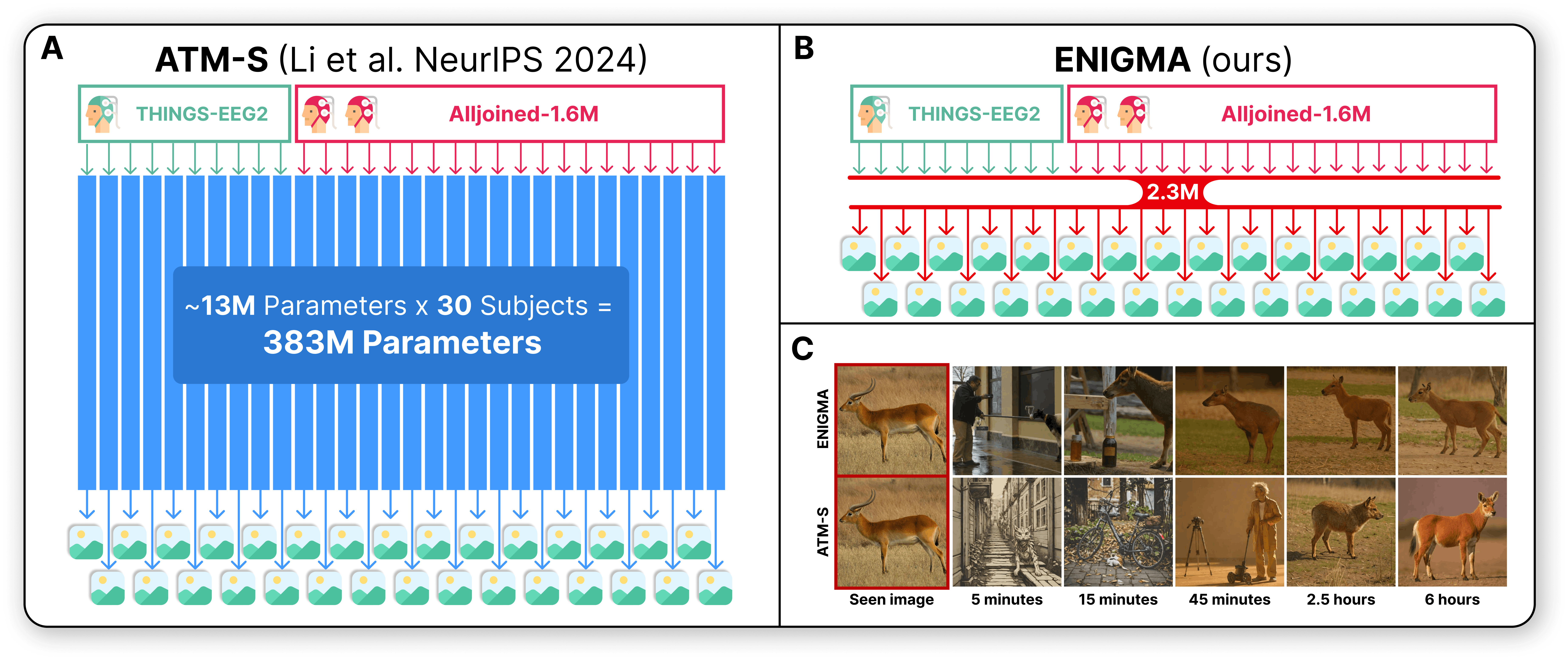}
\captionof{figure}{\textbf{A}: ATM-S \cite{Li2024} vs \textbf{B}: \textbf{ENIGMA} (ours) comparison of model size and multi-subject capabilities on THINGS-EEG2 \cite{Gifford2022} (green cap, 10 subjects) and Alljoined-1.6M \cite{Alljoined-1.6M} (red cap, 20 subjects) datasets. \textbf{C}: Comparison of \textbf{ENIGMA} and ATM-S across training data scale.}
\vspace{5pt}
\label{figure:header}
\end{strip}

\begin{abstract}
To be practical for real-life applications, models for brain-computer interfaces must be easily and quickly deployable on new subjects, effective on affordable scanning hardware, and small enough to run locally on accessible computing resources. To directly address these current limitations, we introduce \textbf{ENIGMA}, a multi-subject electroencephalography (EEG)-to-Image decoding model that reconstructs seen images from EEG recordings and achieves state-of-the-art (SOTA) performance on the research-grade THINGS-EEG2 and consumer-grade AllJoined-1.6M benchmarks, while fine-tuning effectively on new subjects with as little as 15 minutes of data. \textbf{ENIGMA} boasts a simpler architecture and requires less than 1\% of the trainable parameters necessary for previous approaches. Our approach integrates a subject-unified spatio-temporal backbone along with a set of multi-subject latent alignment layers and an MLP projector to map raw EEG signals to a rich visual latent space. We evaluate our approach using a broad suite of image reconstruction metrics that have been standardized in the adjacent field of fMRI-to-Image research, and we describe the first EEG-to-Image study to conduct extensive behavioral evaluations of our reconstructions using human raters. Our simple and robust architecture provides a significant performance boost across both research-grade and consumer-grade EEG hardware, and a substantial improvement in fine-tuning efficiency and inference cost. Finally, we provide extensive ablations to determine the architectural choices most responsible for our performance gains in both single and multi-subject cases across multiple benchmark datasets. Collectively, our work provides a substantial step towards the development of practical brain-computer interface applications.

\end{abstract}
\vspace{-20pt}

\section{Introduction}
\label{intro}
Reconstructing visual experiences from brain activity has long been a goal of both neuroscience and machine learning, and a foundational step for building decoding algorithms for practical brain-computer interface (BCI) applications targeting states like mental images \cite{NSDImagery}. Decoding visual information encoded in human brain activity could help researchers better understand cognitive processes, and could also be useful in a clinical setting \cite{pearson2015mental}, where millions of patients are left unable to communicate through conventional means as a result of traumatic brain injuries, and many common afflictions manifest as a profound dysregulation of unwanted or confusing visual experiences \cite{holmes2010mental}. While the Natural Scenes Dataset (NSD) \cite{allen_massive_2022, NSDImagery} has yielded striking functional magnetic resonance imaging (fMRI)-based reconstructions of seen images using latent diffusion models \cite{Scotti2024MindEye2}, electroencephalography (EEG)-based reconstruction remains challenging due to EEG's low signal-to-noise ratio and spatial resolution. Despite these limitations, EEG remains appealing for real-time BCI applications because of its temporal precision and inexpensive, portable form factor.

Existing EEG-to-Image decoding research spans a wide range of architectural approaches: \citet{Fei2024-perceptogram} demonstrates a simple linear mapping from EEG to an expressive CLIP (Contrastive Language-Image Pretraining \cite{radford_learning_2021}) image embedding space combined with a pre-trained diffusion model, while \citet{Li2024} proposes a more complex architecture (ATM-S) utilizing a transformer-based brain encoder and a two-stage generation process utilizing the diffusion prior introduced in \citet{scotti_reconstructing_2023}. These approaches have laid important groundwork for this field of research, however, current EEG-to-Image models face three critical barriers that prevent their deployment in practical BCI applications.

\textbf{Rapid adaptation to new subjects remains an unsolved challenge.} All existing methods for EEG-to-Image decoding \cite{Song2024, Fei2024-perceptogram, Li2024} require training specialized models from scratch for each subject using hours of training data, which is impractical for real-world BCI applications that demand rapid functionality on new subjects. While recent work in fMRI-to-Image has made progress on this front—\citet{Scotti2024MindEye2} reduced calibration requirements from 40 hours to 1 hour through efficient fine-tuning—even this breakthrough remains insufficient for practical deployment, as it still requires both an expensive 7 tesla fMRI scanner and a full hour of subject-specific data collection. Critically, no EEG-to-Image model has attempted to enable rapid fine-tuning on new subjects, despite EEG's advantages in accessibility and portability. We hypothesize that this gap has persisted because of EEG's noisy spatio-temporal signal characteristics, making it difficult to leverage generalizable knowledge learned from other subjects while quickly adapting to the idiosyncratic neural patterns of a new individual.

\textbf{Current approaches lack robustness when applied to affordable hardware.} The recent release of the Alljoined-1.6M dataset—designed for evaluating EEG-to-Image models on consumer-grade hardware—revealed a troubling pattern: many recent models, especially complex architectures such as ATM-S, are not robust to drops in hardware quality and fail at a much higher rate when deployed in noisier recording environments \cite{Alljoined-1.6M}. This brittleness represents a fundamental obstacle to democratizing BCI technology, as the vast majority of potential users and downstream applications cannot access research-grade EEG equipment. The challenge lies in learning representations that capture the underlying neural encoding of visual information in a way that generalizes across varying levels of signal quality, electrode density, and noise characteristics. Without addressing this weakness, EEG-to-Image models will remain confined to laboratory settings, unable to translate to the accessible, consumer-grade hardware that many practical BCI applications require.

\textbf{Existing architectures are too computationally prohibitive for large-scale deployment.} Most EEG-to-Image models are quite large relative to the amount of data they are trained on, and require separate models to be trained for each subject, resulting in a linear increase in what is already a very large functional model size as models are deployed across multiple users \cite{Song2024, Fei2024-perceptogram}. While \citet{Li2024} demonstrated that a unified model can be trained across multiple subjects, doing so without any mechanism for handling subject-specific differences led to a substantial performance drop; attaining reasonable performance still required training separate models for each subject. These architectural and size limitations constitute a stubborn barrier for widespread deployment: running these large models on edge devices or in settings without access to server-grade GPUs is almost impossible, and supporting inference on multiple subjects requires scaling computational resources proportionally. The core difficulty lies in designing a model architecture with performance \textit{and} efficiency in mind, and in ensuring that a given approach can maintain subject-specific decoding performance while sharing the vast majority of parameters across subjects.

In this paper, we introduce \textbf{ENIGMA} (\textbf{E}EG \textbf{N}eural \textbf{I}mage \textbf{G}enerator for \textbf{M}ulti-subject \textbf{A}pplications), a multi-subject model for reconstructing seen images from EEG data that directly addresses each of these three critical weaknesses. Our work includes several notable contributions:

\textbf{(1)} \textbf{ENIGMA} is the first EEG-to-Image model to quickly fine-tune on new subjects with as little as 15 minutes of data, making it effective for practical downstream use cases (Figure \ref{figure:header}C). \textbf{(2)} Our model achieves robust performance across hardware quality levels, producing SOTA performance on both available benchmark datasets, and a substantial performance gain on Alljoined-1.6M, collected onconsumer-grade hardware. \textbf{(3)} \textbf{ENIGMA} is unified across subjects and datasets, requiring less than 1\% of the trainable parameters to decode the 30 subjects across THINGS-EEG2 and Alljoined-1.6M datasets ($5.5$x reduction vs ATM-S $\times$ 30 subjects = $165$x reduction in total parameters) when compared to single subject approaches (Figure \ref{figure:header}AB). \textbf{(4)} We are the first EEG-to-Image work to provide extensive evaluations using behavioral experiments with human raters, a standard in the adjacent field of fMRI-to-Image research, and we conduct a detailed ablation analysis investigating which architectural aspects are most effective across differences in hardware quality and multi-subject configurations.

Our work demonstrates significant progress towards overcoming the three fundamental barriers that have prevented EEG-to-Image decoding from practical deployment. By enabling rapid fine-tuning with minimal data, maintaining robust performance across hardware configurations, and achieving dramatic improvements in parameter efficiency, \textbf{ENIGMA} represents a sizable step on the path to bring real-time, accessible "mind-to-image" translation from laboratory demonstration to practical clinical, consumer, and research applications.

\vspace{-5pt}
\section{Related Work}
\vspace{-5pt}

\textbf{fMRI-to-Image Reconstruction.} The advent of generative diffusion models has revolutionized neural decoding for adjacent scanning modalities like fMRI \cite{ozcelik2023braindiffuser, scotti_reconstructing_2023,Scotti2024MindEye2,takagi2022_decoding,takagi2023improving, kneeland_brain-optimized_2023, kneeland2023reconstructing, kneeland_second_2023}. \citet{takagi2022_decoding} demonstrated some of the first high-resolution image reconstructions from fMRI by mapping fMRI activity into the latent space of a diffusion model. \citet{ozcelik2023braindiffuser} were the first to show that latent diffusion can reconstruct natural scenes from fMRI with high semantic fidelity, and defined a set of evaluation metrics combining low-level (pixel-wise) and high-level (feature-based) similarity measures that have become standard \cite{ozcelik2023braindiffuser, scotti_reconstructing_2023,Scotti2024MindEye2,takagi2022_decoding,takagi2023improving, kneeland_brain-optimized_2023, kneeland2023reconstructing, kneeland_second_2023}. While fMRI enables finer-grained reconstructions due to its high spatial resolution, EEG’s superior temporal resolution and portability makes it more suited for real-time applications, despite its lower signal fidelity. 

\textbf{Multi-Subject fMRI Models.} Recent work in fMRI-to-Image decoding has demonstrated the effectiveness of multi-subject pretraining for rapid adaptation to new subjects. \citet{Scotti2024MindEye2} introduced MindEye2, which employs a shared-subject functional alignment approach where subject-specific ridge regression linearly maps each individual's fMRI voxel patterns to a common 4096-dimensional latent space, followed by a shared non-linear pipeline across all subjects. By pretraining on 7 subjects and fine-tuning on new subjects, MindEye2 achieves high-quality reconstructions with as little as 1 hour of calibration data—a 40x reduction compared to single-subject approaches. However, such linear alignment strategies on the input data do not translate well to EEG-to-Image decoding due to fundamental differences in signal characteristics: EEG exhibits complex spatio-temporal dynamics with highly variable signal topology across subjects, lower spatial resolution, and substantially higher noise levels compared to the relatively stable spatial patterns of fMRI voxel activity. These challenges necessitate different architectural approaches for achieving robust multi-subject EEG decoding, which we address in this work.

\textbf{EEG-based Visual Decoding.} Decoding visual content from EEG recordings uses a wide array of approaches spanning classification, retrieval, and image reconstruction. While early studies collected EEG recordings in response to visual stimuli \cite{Spampinato2017}, many contained confounds that made it difficult to decode true semantic content \cite{Li2020}, highlighting a need for better data and methods. This critique, along with improvements in EEG preprocessing and experimental design, led to the development of new datasets as part of the THINGS initiative. \citet{Grootswagers2022} released THINGS-EEG (50 subjects, 1,854 concepts) using rapid serial visual presentation, and \citet{Gifford2022} further improved upon THINGS-EEG with THINGS-EEG2, emphasizing trial randomization and quality control, which has since become a standard benchmark for EEG vision decoding. On THINGS-EEG2, new methods \cite{Song2024, Li2024} employing contrastive learning between image and EEG features have achieved significant gains in zero-shot object classification. Alljoined-1.6M \cite{Alljoined-1.6M} extends this paradigm to twice as many subjects and to a consumer-grade EEG hardware setup, providing a new set of tools for developing and evaluating EEG-based visual decoding methods for practical use.

\textbf{EEG-to-Image Reconstruction.} In the space of EEG-to-Image reconstruction methods, \citet{Li2024} introduced a specialized EEG encoder called the Adaptive Thinking Mapper (ATM-S), which uses a two-stage decoding approach comprising a transformer, a CNN, an MLP, and a diffusion prior before using the decoded CLIP vector to generate an image reconstruction using a diffusion model. The complexity of the ATM-S architecture requires careful tuning of many intricate architectural components and multiple sequential training stages to be successful. It was also shown with the release of the Alljoined-1.6M dataset that this degree of complexity results in brittle performance on lower grade EEG hardware \cite{Alljoined-1.6M}. While the architecture does support multi-subject training through a learned subject embedding in the transformer stage, training ATM-S on multiple subjects produces a substantial performance drop, so for most analyses in this work we use the model in its single-subject configuration.

Inspired by \citet{ozcelik2023braindiffuser}, Perceptogram \citep{Fei2024-perceptogram} utilizes a linear transform to map EEG recordings to a CLIP embedding space, and generates images directly from the predicted embeddings using a diffusion model. While its reconstructions still contain less detail than fMRI-based reconstruction, the approach of \citep{Fei2024-perceptogram} demonstrates the significant power of robust linear models in producing recognizable images from low SNR brain activity patterns.

\begin{figure*}[!htb]
\centering
\vspace{-15pt}
\includegraphics[width=\textwidth]{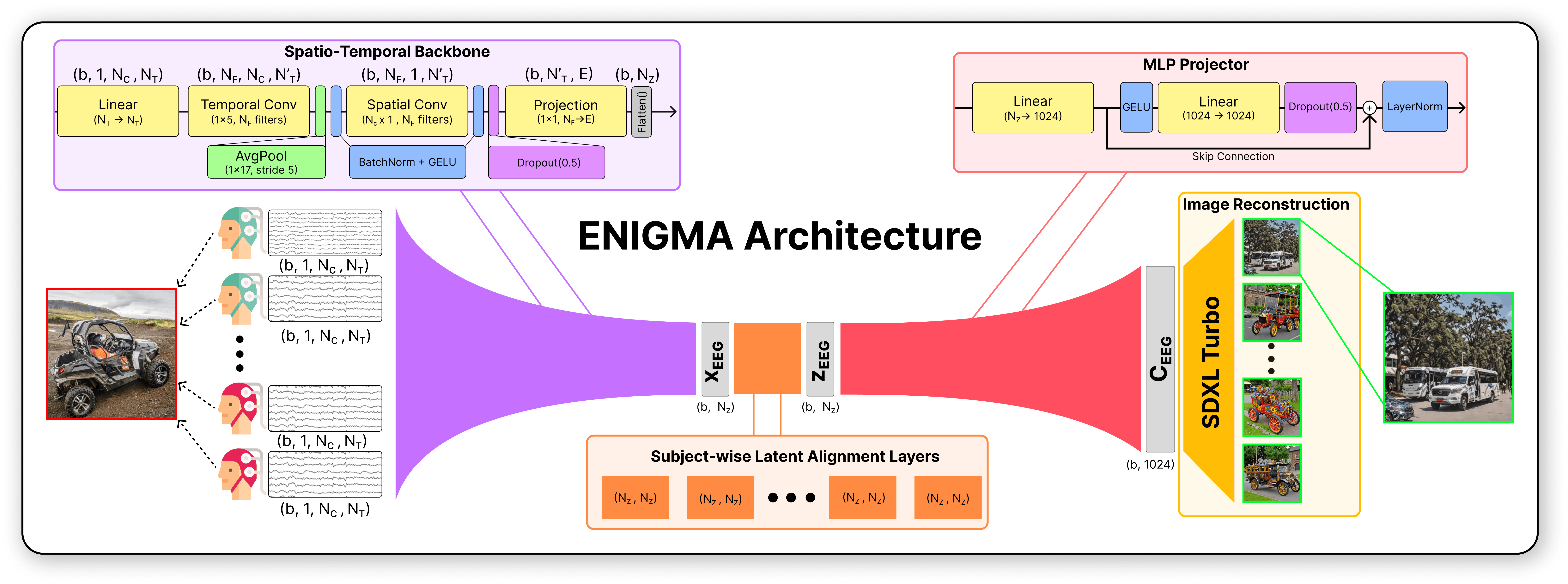}
\vspace{-25pt}
\caption{During training, brain activity from each subject is passed through a shared pathway of spatio-temporal convolutions, producing an intermediate latent vector $x_{\textrm{EEG}}$ from all subjects, this is then passed through a set of subject-specific latent alignment layers to produce an aligned latent embedding $z_{\textrm{EEG}}$. This latent is passed through a fully connected MLP projection layer to produce the output $c_{\textrm{EEG}}$ vector, which is reconstructed into an image using SDXL. Details of these procedures are provided in Section \ref{architecture}.}
\label{figure:architecture}
\vspace{-10pt}
\end{figure*}

\section{\textbf{ENIGMA}}

\subsection{Methodology}
We designed our model to adhere to several key requirements for practical BCI applications: \textbf{(1)} high performance on both research-grade and consumer-accessible EEG hardware, \textbf{(2)} a subject-unified model architecture that rapidly fine-tunes on new subjects, and \textbf{(3)} a small, scalable, and efficient design that minimizes model complexity and compute requirements for both training and inference.

To meet these requirements, our proposed model, \textbf{ENIGMA}, has 4 components: \textbf{(1)} a spatio-temporal convolutional neural network to learn a set of robust features from the spatial and temporal dimensions of the input signal, \textbf{(2)} a set of subject-wise latent alignment layers to capture and navigate subject-specific differences in the latent space of the model, \textbf{(3)} an MLP projector to the ViT-H/14 CLIP \cite{radford_learning_2021} embedding space, and \textbf{(4)} a pretrained diffusion model (i.e. SDXL Turbo) that can invert a CLIP embedding into an image.

Unlike previous approaches, our model can operate in three distinct modes: single-subject (trained on one subject at a time), multi-subject (trained on multiple subjects in parallel using shared model weights), and fine-tuned (pretrained on one or more subjects, and fine-tuned on a target subject). In practice \textbf{ENIGMA} has near identical performance in the single-subject and fine-tuned conditions when using all available training data (see Figure \ref{fig:finetune}), and so we primarily report performance metrics in the single and multi-subject configurations, while examining the fine-tuning efficiency benefits provided by our pretraining approach.

\subsection{Datasets}
\label{sec:datasets}
\textbf{THINGS-EEG2} \citep{Gifford2022} is the current public benchmark for EEG-based visual decoding. It contains 64-channel ActiChamp (costing \(\sim\)\$60,000) recordings recorded at 1000Hz from 10 participants viewing 16740 unique images in a rapid serial visual-presentation paradigm. 200 of these images are designated as the testing set and are repeated 80 times each, while the remaining 16540 images in the training set are repeated 4 times each, for a total of \(\sim\)820k trials across the whole dataset.

\textbf{Alljoined-1.6M} \cite{Alljoined-1.6M} is a follow-up corpus of EEG responses to visual stimuli collected with a much cheaper 32-channel Emotiv Flex 2 gel headset (\(\sim\)\$2,200) at 250Hz comprising the same stimuli and experimental paradigm as THINGS-EEG2, across 20 new subjects, for a total of \(\sim\)1.6M trials.

Our preprocessing steps are in Appendix \ref{app:data_proc}. When reproducing other methods, we follow their preparation and preprocessing steps. %

\subsection{Architecture}
\label{architecture}
The following components of our \textbf{ENIGMA} architecture are depicted in Figure~\ref{figure:architecture}.

\textbf{Spatio-Temporal Backbone} The preprocessed EEG data is passed through an embedding module that applies convolutions over the temporal and spatial dimensions of our data. We treat multichannel EEG with $N_C$ channels and $N_T$ time points as an ``image" of shape $1 \times N_C \times N_T$. 

A temporal 2D convolution (kernel $(1,5)$, $N_F=40$ feature maps) is followed by average pooling over time (kernel $(1,17)$, stride $5$), batch normalization (BN), and a GELU non-linearity:
\vspace{-5pt}
\begin{align*}
z_1 &= \text{Conv2D}_{1\times 5}^{N_F}(EEG)\\
z_2 &= \text{AvgPool}_{1\times 17,5}(z_1)\\
h_1 &= \text{GELU}(\text{BN}(z_2))
\end{align*}
where $h_1 \in \mathbb{R}^{N_F\times N_C\times N_T'}$ with $N_T' = \left\lfloor\frac{N_T-21}{5}\right\rfloor+1$ (i.e., $N_T=250 \Rightarrow N_T'=46$). Next, a spatial convolution (kernel $(N_C,1)$, $N_F$) integrates information across channels, followed by BN and GELU:
\begin{align*}
z_3 &= \text{Conv2D}_{N_C\times1}^{N_F}(h_1)\\
h_2 &= \text{GELU}(\text{BN}(z_3))
\end{align*}
where $h_2\in\mathbb{R}^{N_F\times1\times N_T'}$.

We apply dropout~\cite{srivastava2014dropout} ($p=0.5$) for regularization and project the features to embedding dimension $E_p=4$ with a $1\times1$ convolution. Finally, we flatten the output $\in \mathbb{R}^{E\times1\times N_T'}$ to obtain a latent embedding vector $X_{EEG}\in\mathbb{R}^{N_z}$ of dimension $N_z = 184$.

\textbf{Subject-wise Latent Alignment Layers} To account for systematic differences between subjects, we learn a set of subject-specific fully-connected linear alignment layers with weights $W_s \in \mathbb{R}^{N_z \times N_z}$ that output an aligned latent representation $z_{EEG}$ across subjects. This alignment module allows downstream modules to learn a unified mapping to the CLIP embedding space across subjects, and is an important piece in allowing the vast majority of model parameters to be shared across multiple subjects with diverse signal dynamics.

\textbf{MLP Projector} After obtaining the aligned $z_{\textrm{EEG}}$ latent embedding, we use a projection head to map it to the final CLIP ViT-H/14 latent dimension $D = 1024$. The head is a feed-forward MLP network with a skip connection: a linear layer from $184$ to $1024$, followed by GELU and dropout, then another linear layer, and finally layer normalization. The residual is added to the output of the second linear layer before normalization, to help stabilize training and allow the model to refine the initial linear projection with non-linear adjustments. The module outputs an EEG-predicted CLIP embedding $c_{\text{EEG}} \in \mathbb{R}^{1024}$.

\textbf{Image Reconstruction}
To generate images from the EEG embedding $c_{\text{EEG}}$, we leverage Stable Diffusion XL Turbo (SDXL) \cite{sauer2024adversarial}, and its associated CLIP ViT-H/14 image-prompt adapter (IP-Adapter) \cite{Ye2023}. The IP-Adapter is a lightweight module inserted into SDXL's cross-attention layers, which enables an image embedding to steer image generation alongside an optional text prompt. We optimize our model to predict the CLIP ViT-H/14 image embeddings expected by SDXL Turbo's IP-adapter as input. Formally, SDXL solves:
\[ x_{T} \sim \mathcal{N}(0,I), \]
\[ x_{t-1} = f_\theta(x_t, c_\text{text}, c_\text{EEG}, t) + \text{noise} \]
for $t=T, T-1, \dots, 0$, where $c_\text{text}$ is the text context (which in our case is an unconditional placeholder embedding) and $c_\text{EEG}$ is our injected EEG image embedding. We run the diffusion for 4 inference steps, which is standard for this version of SDXL.

\textbf{Loss Functions and Training} 
\label{loss}
Following \citet{Li2024}, we align the EEG embedding $c_{\text{EEG}}$ to the CLIP ViT-H/14 image embedding of the stimulus image $f_{\text{CLIP}}(\text{image}) \in \mathbb{R}^{1024}$ by minimizing the Mean-Squared Error (MSE) between the two and regularizing with the InfoNCE contrastive loss \cite{InfoNCE, radford_learning_2021}. The former matches the EEG embedding to its corresponding image embedding in CLIP latent space, and the latter ensures that the embedding retains relevant directional semantics within the CLIP manifold, while learning to discard the subject and session-specific information. The relative weight of these two losses is modulated by $\lambda=0.5$, our overall loss function can be seen in Equation \ref{eq:loss}, and the full mathematical formulations of our loss functions can be found in Appendix \ref{app:loss}. %

\begin{equation}
\label{eq:loss}
\begin{split}
\mathcal{L} &= \text{MSE}(c_{EEG}, f_{CLIP}(\text{image})) \\
&\quad + \lambda \cdot \text{InfoNCE}(c_{EEG}, \text{norm}(f_{CLIP}(\text{image})))
\end{split}
\end{equation}

\textbf{ENIGMA} was trained in FP32 on an RTX 3090 GPU using the AdamW optimizer with learning rate 3e-4 and batch size 512. In the multi-subject configuration, we train for 150 epochs on the training split of both THINGS-EEG2 and Alljoined-1.6M simultaneously (30 subjects, \(\sim\)2M data points averaged for each image to \(\sim\)500k training trials). In this configuration, the model takes 5.5 hours to train across all 30 subjects, and 10 minutes to train on a single subject. We note that in practice our model could also be trained on GPUs with as little as 8GB of VRAM and fine-tuned in as little as 2 minutes when using a 15 minute calibration period. Such minimal computational requirements allow \textbf{ENIGMA} to be rapidly deployable on edge computing devices, facilitating a wide array of downstream applications.

\section{Results}
\label{sec:analysis}
\begin{figure*}[!htb]
\centering
\includegraphics[width=\textwidth]{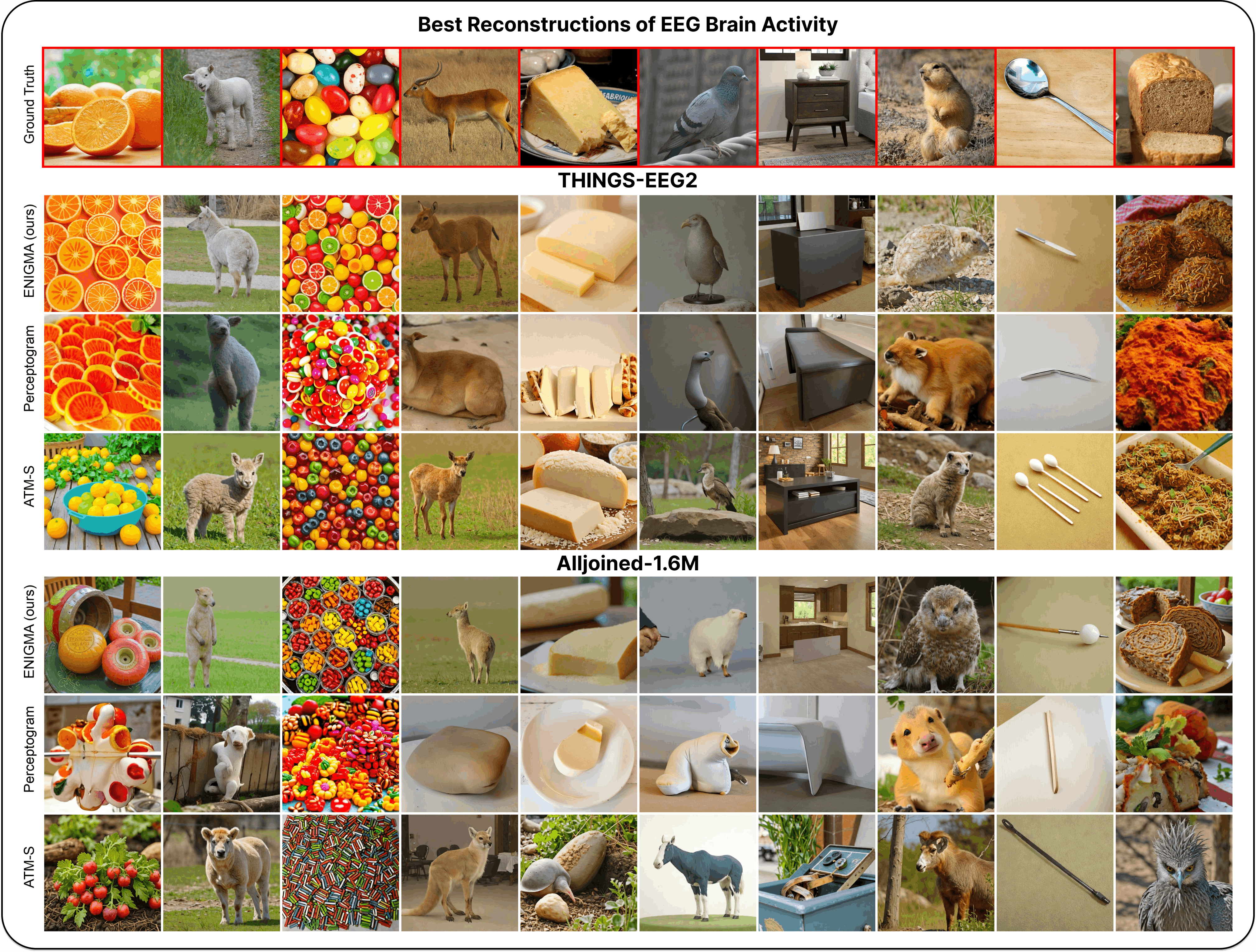}
\vspace{-13pt}
\caption{Qualitative comparison of reconstruction methods on seen stimuli from THINGS-EEG2 and Alljoined-1.6M. Reconstructions selected are the outputs sampled from each method and stimulus with the highest scores on all of the image feature metrics in Table \ref{table:featuremetrics}.}
\label{figure:best}
\end{figure*}

We report results on two of the most recent and prominent EEG-to-Image benchmarks: THINGS-EEG2 \cite{Gifford2022} and Alljoined-1.6M \cite{Alljoined-1.6M}, and evaluate \textbf{ENIGMA} against the only available EEG-to-Image baselines for those datasets, Perceptogram \cite{Fei2024-perceptogram} and ATM-S \cite{Li2024}. Figure \ref{figure:best} presents a set of the best reconstructed images, comparing our method with available baselines on both available datasets. These examples illustrate typical outcomes: our reconstructions generally capture the correct high-level object, e.g., oranges, sheep, furniture, etc. Perceptograms are usually blurrier and sometimes miss the object entirely, e.g., producing a vague shape or significant visual distortions. The ATM-S images are categorically similar, but are often less visually specific to the precise object being decoded. Examples of median and worst-case reconstructions can also be seen in Appendix \ref{app:median_worst}. %

\clearpage

\begin{table*}[!htb]
    \vspace{-10pt}
    \centering
    \setlength{\tabcolsep}{2pt}
    \small
    \resizebox{\textwidth}{!}{
    \begin{tabular}{lcccccccccccccc}
        \toprule
        Method & \multicolumn{2}{c}{Model Properties} & \multicolumn{2}{c}{Low-Level} & \multicolumn{6}{c}{High-Level} & \multicolumn{3}{c}{Retrieval} & \multicolumn{1}{c}{Human Raters} \\
        \cmidrule(lr){2-3}\cmidrule(lr){4-5}\cmidrule(lr){6-11}\cmidrule(lr){12-14}\cmidrule(l){15-15}
        & \# of Parameters $\downarrow$ & Inference GFLOPS $\downarrow$ & PixCorr $\uparrow$ & SSIM $\uparrow$ & Alex(2) $\uparrow$ & Alex(5) $\uparrow$ & Incep $\uparrow$ & CLIP $\uparrow$ & Eff $\downarrow$ & SwAV $\downarrow$ & Top-1 $\uparrow$ & Top-5 $\uparrow$ & Top-10 $\uparrow$ & Ident.\ Acc.\ $\uparrow$ \\
        \midrule
        \multicolumn{14}{c}{\textbf{THINGS-EEG2 (10 subjects)}} \\
        \midrule
        \textbf{ENIGMA (Multi-Subject)} & \textbf{2,376,842} & \textbf{294.4} & \textbf{0.1668} & \textbf{0.4264} & \textbf{82.99\%} & \textbf{89.12\%} & \textbf{76.54\%} & \textbf{80.33\%} & \textbf{0.8577} & \textbf{0.5399} & \textbf{22.55\%} & \textbf{50.75\%} & \textbf{64.05\%} & \textbf{86.04\%} \\
        ATM-S (Multi-Subject)           & \underline{12,815,311} & \underline{3,858.6} & \underline{0.072} & \underline{0.403} & \underline{57.09\%} & \underline{58.99\%} & \underline{52.86\%} & \underline{55.04\%} & \underline{0.963} & \underline{0.663} & \underline{16.20\%} & \underline{45.10\%} & \underline{62.20\%} & \underline{56.82\%} \\
        \midrule
        \textbf{ENIGMA (Single-Subject)} & \textbf{13,896,820} & \textbf{294.4} & \underline{0.1718} & \underline{0.4233} & \underline{83.64\%} & \textbf{89.49\%} & \textbf{77.65\%} & \textbf{81.48\%} & \textbf{0.8547} & \textbf{0.5403} & \underline{27.60\%} & \underline{59.35\%} & \underline{71.15\%} & \textbf{86.82\%} \\
        ATM-S (Single-Subject)           & \underline{128,153,110} & 3,858.6 & 0.136 & 0.392 & 73.85\% & 80.83\% & 67.56\% & 71.28\% & 0.909 & 0.601 & \textbf{30.15\%} & \textbf{60.15\%} & \textbf{73.60\%} & 77.14\% \\
        Perceptogram (Single-Subject)    & 4,731,924,800 & \underline{2,807.8} & \textbf{0.247} & \textbf{0.431} & \textbf{85.46\%} & \underline{88.03\%} & \underline{70.40\%} & \underline{71.98\%} & \underline{0.902} & \underline{0.581} & -- & -- & -- & \underline{79.17\%} \\
        \midrule
        \multicolumn{14}{c}{\textbf{Alljoined-1.6M (20 subjects)}} \\
        \midrule
        \textbf{ENIGMA (Multi-Subject)} & \textbf{2,376,842} & \textbf{588.8} & \textbf{0.0852} & \textbf{0.4175} & \textbf{68.33\%} & \textbf{73.40\%} & \textbf{63.14\%} & \textbf{66.38\%} & \textbf{0.9259} & \textbf{0.6127} & \textbf{6.00\%} & \textbf{18.85\%} & \textbf{28.80\%} & \textbf{70.74\%} \\
        ATM-S (Multi-Subject)           & \underline{12,765,711} & 7,717.2 & \underline{0.068} & \underline{0.417} & \underline{53.49\%} & \underline{53.36\%} & \underline{50.72\%} & \underline{51.46\%} & \underline{0.965} & \underline{0.668} & \underline{0.73\%} & \underline{4.13\%} & \underline{7.55\%} & \underline{52.18\%} \\
        \midrule
        \textbf{ENIGMA (Single-Subject)} & \textbf{27,793,640} & \textbf{588.8} & 0.0852 & \textbf{0.4175} & \textbf{68.33\%} & \textbf{73.40\%} & \textbf{63.14\%} & \textbf{66.38\%} & \textbf{0.9259} & \textbf{0.6127} & \textbf{6.00\%} & \textbf{16.25\%} & \textbf{25.35\%} & \textbf{71.82\%} \\
        ATM-S (Single-Subject)           & \underline{255,314,220} & 7,717.2 & \underline{0.090} & 0.374 & 55.91\% & 58.25\% & 54.07\% & 56.25\% & 0.960 & 0.673 & \underline{0.50\%} & \underline{2.00\%} & \underline{5.00\%} & 60.31\% \\
        Perceptogram (Single-Subject)    & 9,463,849,600 & \underline{5,615.6} & \textbf{0.094} & \underline{0.401} & \underline{67.36\%} & \underline{69.28\%} & \underline{58.18\%} & \underline{59.94\%} & \underline{0.945} & \underline{0.637} & -- & -- & -- & \underline{62.00\%} \\
        \bottomrule
\end{tabular}
    }
    \caption{Comparison of  EEG-to-Image reconstruction models on the THINGS‑EEG2 and Alljoined-1.6M datasets via image similarity metrics. Parameter counts and GFLOPS are computed by adding up the number of parameters and compute necessary to decode all subjects in each dataset (10 subjects for THINGS-EEG2, 20 for Alljoined-1.6M) in each configuration (single vs multi-subject). Details on the human identification accuracy metric are provided in Section \ref{sec:human_experiments} and Appendix \ref{app:behavioral} For the \# of parameters, GFLOPS, EffNet-B, and SwAV, lower is better. For all other metrics, higher is better. Bold indicates best performance, and underlines second-best performance. Additional details on the metrics are in Appendix \ref{app:metrics}.} %
    \label{table:featuremetrics}
\end{table*}

\clearpage

\subsection{Quantitative Evaluations}

Table \ref{table:featuremetrics} summarizes the quantitative performance of \textbf{ENIGMA} and  baseline methods on both benchmarks in single and multi-subject configurations. For all methods, we output 10 reconstructions per test sample from each method and report averaged image feature metrics across them. For multi-subject configurations (our primary evaluation target), \textbf{ENIGMA} achieves the best scores on all metrics, indicating that the latent alignment layers are enabling our model to align representations across subjects from both datasets. For single-subjects, our model still provides SOTA performance on the majority of metrics. 

\subsection{Human Behavioral Evaluations}
\label{sec:human_experiments}
For brain decoding models to be deployed in BCI applications, their outputs must be meaningfully interpretable to users, scientists, and clinicians. While many of the metrics in Table \ref{table:featuremetrics} are commonly used as proxies for perceptual quality, prior research has established that these automated metrics often fail to align with human assessments of content \cite{peceptualsimilarity} or quality \cite{pickapic}. Human judgment of reconstruction quality is therefore an essential performance metric for evaluating EEG-to-Image models.

To address this, we conducted a carefully controlled, large-scale online behavioral experiment where human raters (n = 545) assessed the quality of reconstructions through 2-alternative forced choice judgments. In each trial, raters determined whether a reconstruction was more similar to its corresponding ground truth image than to a randomly selected reconstruction from the same method, dataset, and subject. This task evaluates whether reconstructions contain meaningful stimulus-specific content—a minimum requirement for practical utility. Detailed experimental protocols are provided in Appendix \ref{app:behavioral}. %

Our results in Table \ref{table:featuremetrics} confirm that \textbf{ENIGMA} achieves a substantial improvement in human identification accuracy across all evaluated conditions, which we believe to be the most reliable indicator of performance on EEG-to-Image tasks. We also confirm that all reconstruction models evaluated here produce above-chance performance in our randomized control trials ($p < 0.001$).

\subsection{Fine Tuning Efficiency}
\label{sec:finetune}

\begin{figure}[!htb]
  \centering
  \includegraphics[width=\columnwidth]{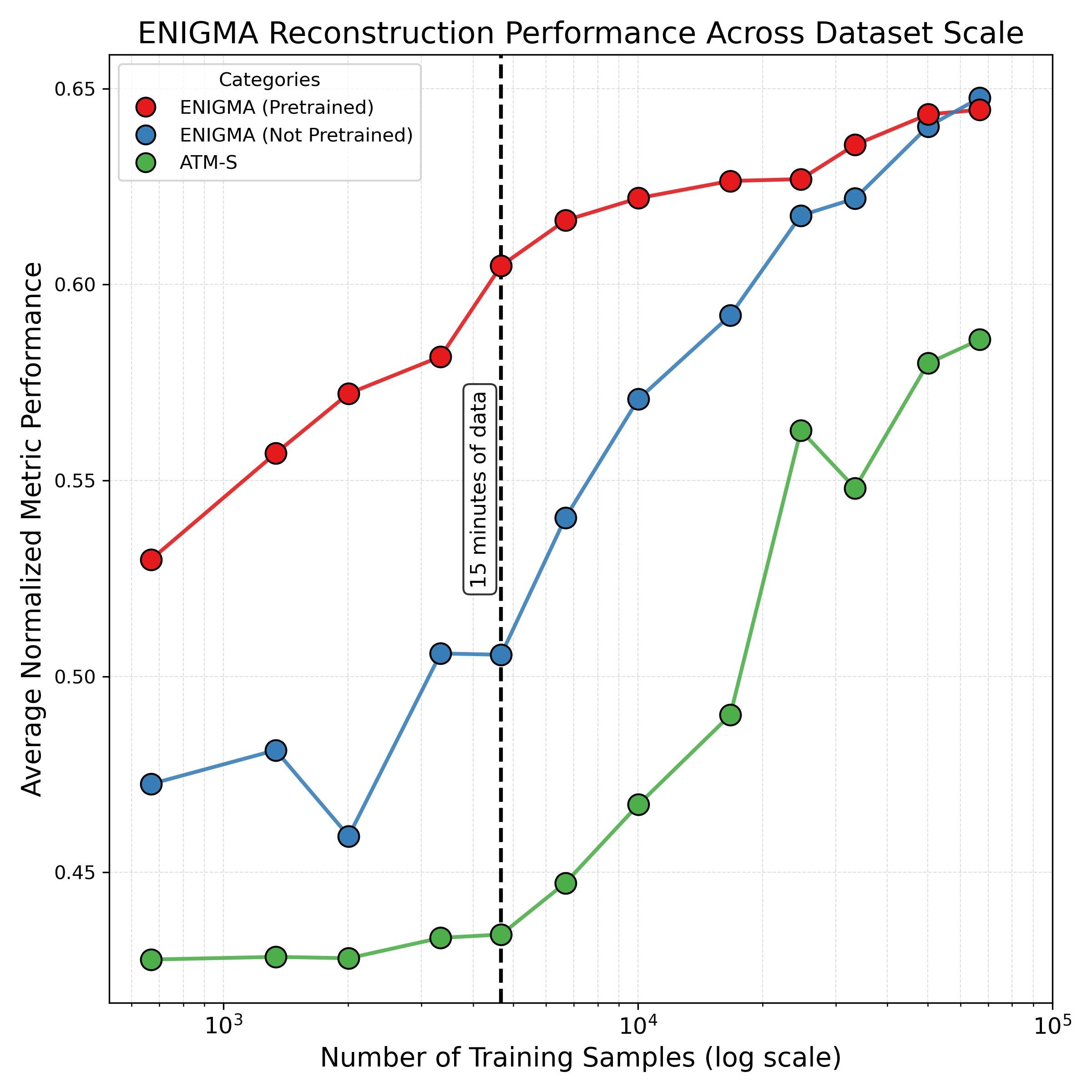}
  \vspace{-15pt}
  \caption{%
    Scaling efficiency of \textbf{ENIGMA} with (red) and without (blue) pretraining on other subjects, and ATM-S \cite{Li2024} (green). Performance is plotted using varying amounts of target subject training/fine-tuning data on a log-scale X-axis. Reconstruction accuracy, evaluated using the normalized average of feature metrics presented in Table \ref{table:featuremetrics}, is plotted on the Y-axis. All metrics are calculated on the median subject (2) of the THINGS-EEG2 dataset.}
  \vspace{-15pt}
  \label{fig:finetune}
\end{figure}

One of the primary goals of our work is to enable high-quality reconstructions from brief calibration sessions suitable for clinical, consumer, and research contexts where extended data collection is infeasible. Figure \ref{fig:finetune} demonstrates this capability by comparing reconstruction performance across \textbf{ENIGMA} models trained or fine-tuned on varying amounts of subject-specific data. The vertical dashed line marks a 15-minute calibration threshold—a reasonable constraint for real-world BCI deployment.
While both pretrained and non-pretrained models improve consistently as training data increases, multi-subject pretraining provides substantial advantages in the low-data regime. At the 15-minute threshold (approximately 4,000 training samples), pretrained \textbf{ENIGMA} surpasses the fully-trained ATM-S architecture, while the non-pretrained baseline fails to produce identifiable reconstructions. Further analysis of the scaling performance on each benchmark dataset and of the number of EEG channels can also be found in Appendices \ref{app:scaling} and \ref{app:channelcount}.These results provide empirical evidence that \textbf{ENIGMA} and its multi-subject latent alignment procedure enable effective transfer learning in EEG-to-Image decoding, addressing a critical gap where no prior work has demonstrated functional performance on a novel subject with such limited calibration data. 

\subsection{Ablation Study}

\begin{figure}[!htb]
  \centering
  \includegraphics[width=\columnwidth]{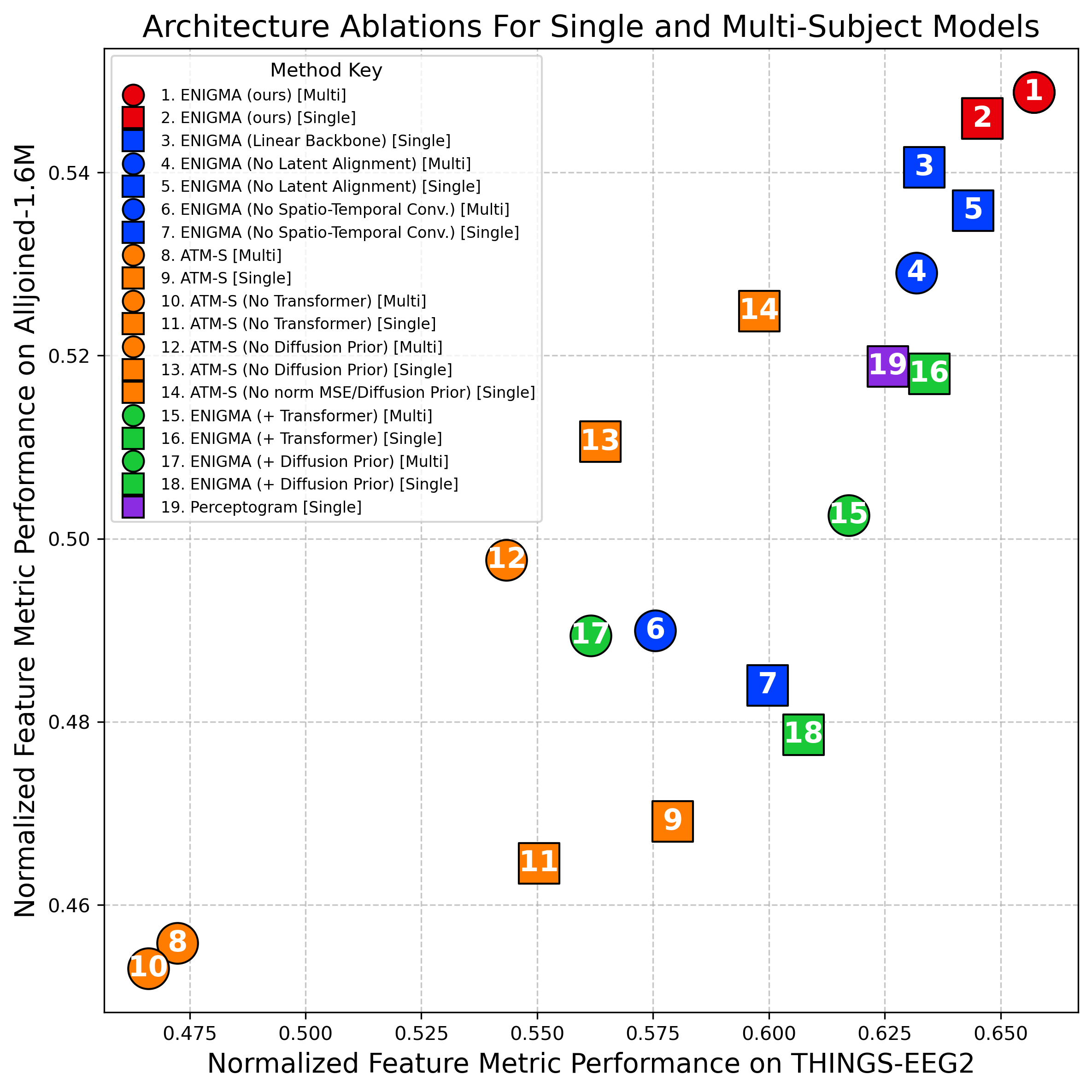}
  \vspace{-15pt}
  \caption{%
    Ablation analyses: model variants (numbered icons) in single (square) and multi-subject (circle) configurations under each ablation type (color) are assessed via the normalized average of all feature metrics (Table \ref{table:featuremetrics}), with THINGS-EEG2 performance on the x-axis and Alljoined-1.6M performance on the y-axis.}
  \vspace{-18pt}
  \label{fig:ablation}
\end{figure}

To rigorously evaluate EEG-to-Image architectures and critically examine why our method succeeds in multi-subject and consumer-grade contexts where ATM-S breaks down, we conducted ablations on key design choices in our method. Colored numeric identifiers refer to the ablation results in Figure \ref{fig:ablation}. \textbf{ENIGMA} is displayed as \textcolor{final}{(1)}, \textcolor{final}{[2]}.

\label{sec: ablations}

\textbf{ENIGMA Modules.} While linear models do not provide any of the parameter efficiency or fine-tuning benefits of our \textbf{ENIGMA} architecture, we find using a well regularized linear backbone \textcolor{enigma}{[3]} in single-subject contexts remains an effective way to decode semantic information from brain activity \cite{ozcelik2023braindiffuser, Fei2024-perceptogram}. We also find that eliminating the latent alignment layer \textcolor{enigma}{(4)}, \textcolor{enigma}{[5]} harms performance disproportionately in multi‑subject contexts, showing that the module specifically helps drive cross‑participant generalization. Removing the spatio‑temporal convolution stack decreases accuracy in all contexts \textcolor{enigma}{(6)}, \textcolor{enigma}{[7]}, confirming that joint space‑time feature extraction is essential for capturing the semantic information encoded in EEG brain activity.

\textbf{ATM‑S Modules.} While ATM‑S’s \textcolor{atm}{(8)}, \textcolor{atm}{[9]} diffusion prior slightly improves its performance on THINGS-EEG2 \textcolor{atm}{[13]}, we find in our analysis that removing it significantly improves performance on data using the consumer-grade EEG hardware in Alljoined-1.6M, and on both datasets in multi-subject contexts \textcolor{atm}{(12)}. We also noticed that if we remove the normalization of the target vectors in the MSE loss computation (see Appendix \ref{app:loss} for details), results improve even further and negate the need for the diffusion prior entirely \textcolor{atm}{[14]}. We observe similar patterns with the transformer encoder \textcolor{atm}{(10)}, \textcolor{atm}{[11]}, which only benefits narrow single-subject THINGS-EEG2 settings. These results suggest that these specific architectural complexities hinder both robustness to lower-SNR data and the ability to capture subject-specific variations in unified multi-subject models, informing our simple and robust approach. %

\textbf{ENIGMA + ATM-S Modules.} To evaluate whether the architectural modules introduced by \citet{Li2024} would be beneficial to \textbf{ENIGMA}, we grafted ATM‑S’s transformer-based encoder and diffusion prior training stage onto \textbf{ENIGMA} \textcolor{enigma+atm}{(15)},\textcolor{enigma+atm}{[16]},\textcolor{enigma+atm}{(17)},\textcolor{enigma+atm}{[18]}. We find that both of these modules harm performance on both datasets in both single and multi-subject contexts, highlighting how our simple and robust design captures all of the necessary information encoded in brain activity without needing additional computationally expensive modules.

\section{Discussion}
\label{discussion}
Here we introduce \textbf{ENIGMA}, a multi-subject EEG-to-Image reconstruction model that directly addresses three critical barriers preventing practical deployment of visual decoding systems: the inability to rapidly fine-tune on new subjects, brittle performance on consumer-grade hardware, and computational prohibitiveness at scale. Through multi-subject latent alignment and a suite of other techniques, \textbf{ENIGMA} achieves performant reconstructions with only 15 minutes of calibration data, an order of magnitude reduction compared to existing approaches that require hours of subject-specific training. Our lightweight spatio-temporal CNN architecture maintains robust performance across both research-grade (THINGS-EEG2) and consumer-grade (Alljoined-1.6M) hardware, demonstrating that architectural simplicity and multi-subject pretraining enable effective performance across signals of varying quality. Finally, by sharing model weights across subjects with only lightweight subject-specific latent alignment layers, \textbf{ENIGMA} achieves a $\sim$165x reduction in parameters for multi-subject deployments, enabling efficient simultaneous inference for multiple users with less than 1\% of the model size.

Motivated by the overlap between vision and mental imagery \cite{BREEDLOVE20202211}, we believe models like \textbf{ENIGMA} that target semantic content encoded in brain activity from visual perception are a meaningful step to more generalized breakthroughs in "mind reading" technology, and look forward to future research building on our work towards these goals.

\subsection{Current Limitations}
\label{limitations}
Despite training on up to 30 participants, \textbf{ENIGMA’s} multi‑subject training and fine-tuning approaches do not introduce any measurable performance gains, i.e., adding more subjects does not meaningfully change the performance ceiling of the model. This aligns with previous findings studying the scaling properties of neuroimaging data \cite{banville2025scaling}, but nonetheless represents an important limitation for the field to overcome. Our model has so far been validated only in a tightly constrained image‑reconstruction paradigm, leaving its utility for more open‑ended BCI tasks untested. We plan to explore the above research avenues in future work.

\section*{Broader Impacts}
\label{broaderimpacts}
While fMRI has dominated neural image reconstruction research due to its superior spatial resolution and downstream decoding performance, its limited accessibility, high cost, and restriction to laboratory settings preclude real-world BCI deployment. \textbf{ENIGMA} demonstrates that EEG, despite its lower signal quality, provides sufficient information for semantically meaningful reconstruction when combined with appropriate architectural choices and training strategies. The convergence of rapid adaptation, hardware robustness, and computational efficiency represents a fundamental shift in the practical viability of EEG-based visual decoding, establishing a foundation for deployable brain-computer interfaces in assistive communication, clinical assessment, and real-time neural decoding applications.

\subsection*{Ethical Considerations}
\label{ethics}
Research aimed at decoding cognitive states is rapidly growing in scope and capability. While these endeavors promise clear downstream benefits, they also raise serious questions about broader societal implications and their potential for misuse. Because of these risks, we stress the importance of developing an ethical framework for the application of brain decoding devices that rigorously safeguards users' data and ensures that the technology is deployed transparently, responsibly, and for the benefit of humankind \cite{sethics}.

\newpage
\appendix
\onecolumn
\section{Appendix}
\subsection{Data Processing and Format} 
\label{app:data_proc}
Raw EEG was stored in standard \texttt{.edf} files and pre-processed with MNE-Python \citep{Gramfort2013}. The Emotiv firmware applies a dual 50/60 Hz notch by default, effectively attenuating frequencies above 43 Hz, so we added a 0.5 Hz high-pass and an extra 60 Hz notch to suppress residual line noise. We epoched continuous recordings from $-200$ ms to 1000 ms relative to image onset. Synchronisation glitches in the Emotiv trigger stream led us to discard 0.55–1.12\% of trials, which was comparable to exclusion rates reported in earlier Emotiv evaluations \citep{badcock2013validation,williams2020validation}. Epochs were then baseline-corrected to the pre-stimulus window and resampled to 250 Hz to match the ATM-S benchmark \citep{Li2024}. Finally, we performed multivariate noise normalization \citep{Guggenmos2018} where we whiten input data to improve signal-to-noise ratio (SNR). Note that we estimate the whitening matrix only on the training partition to avoid data contamination. This yielded samples $x_\textrm{EEG} \in \mathbb{R}^{N_C \times N_T}$ with $N_C=64$ channels for THINGS-EEG2, $N_C=32$ channels for Alljoined-1.6M, and $N_T=250$ time points for both datasets. For multi-subject models, all models require the same channel count across all subjects, and so for these models we subsample THINGS-EEG2 to the same $32$ channels present in Alljoined-1.6M. For an analysis of this step on performance, see Appendix \ref{app:channelcount}. Both the THINGS-EEG2 and Alljoined-1.6M datasets contain 4 image repetitions per training sample, and 80 repetitions per inference sample. For training and inference, we average together these multiple trials for each image presentation to further boost the SNR of the data. 

\subsection{Loss function formulations}
\label{app:loss}
For the loss functions behind the training of ENIGMA, we use common, well-defined loss functions for mean-squared error and symmetric InfoNCE contrastive loss widely utilized in the literature, with precise mathematical definitions below.

\begin{equation}
L = \text{MSE}(c_{EEG}, f_{CLIP}(\text{image})) + \lambda \cdot \text{InfoNCE}(c_{EEG}, \text{norm}(f_{CLIP}(\text{image})))
\end{equation}

\begin{equation}
\text{MSE} = \frac{1}{N} \sum_{i=1}^{N} \|c_{EEG}^i - f_{CLIP}(\text{image})^i\|_2^2
\end{equation}

\begin{equation}
\text{InfoNCE} = \frac{L_{EEG \to Image} + L_{Image \to EEG}}{2}
\end{equation}

\begin{equation}
L_{EEG \to Image} = -\frac{1}{N} \sum_{i=1}^{N} \log\left[\frac{\exp(c_{EEG}^i \cdot f_{CLIP}(\text{image})^i / \tau)}{\sum_{j=1}^{N} \exp(c_{EEG}^i \cdot f_{CLIP}(\text{image})^j / \tau)}\right]
\end{equation}

\begin{equation}
L_{Image \to EEG} = -\frac{1}{N} \sum_{i=1}^{N} \log\left[\frac{\exp(c_{EEG}^i \cdot f_{CLIP}(\text{image})^i / \tau)}{\sum_{j=1}^{N} \exp(c_{EEG}^j \cdot f_{CLIP}(\text{image})^i / \tau)}\right]
\end{equation}

Here:
\begin{itemize}
    \item $c_{EEG} \in \mathbb{R}^{1024}$ is the output of our EEG encoding pipeline (after MLP projection)
    \item $f_{CLIP}(\text{image}) \in \mathbb{R}^{1024}$ is the CLIP image embedding
    \item $\cdot$ denotes the dot product between normalized embeddings
    \item $\tau = 1/\text{logit\_scale}$ is the learnable temperature parameter
    \item The diagonal elements ($i=j$) represent positive EEG-image pairs
    \item All off-diagonal elements serve as negatives within the batch
\end{itemize}

The symmetric formulation ensures bidirectional alignment between EEG and image modalities. By maximizing $c_{EEG}^i \cdot f_{CLIP}(\text{image})^i$ for paired samples while minimizing similarity with other samples in the batch, we learn a joint embedding space capturing the semantic correspondence between brain activity and visual stimuli.

In our research we also noticed that the MSE loss term of the ATM-S backbone \cite{Li2024} was computed on unit-normalized CLIP targets: $\text{norm}(f_{CLIP}(\text{image}))$, inadvertently optimizing for angular alignment (cosine similarity) rather than Euclidean distance in the embedding space. We chose not to normalize $f_{\text{CLIP}}(\text{image})$ in the MSE component to ensure that the learned $c_{\text{EEG}}$ respects the geometry of the CLIP embedding space, and note that doing so negates the need for the secondary diffusion prior training stage in \citet{Li2024} that was previously necessary to learn the magnitude of $f_{\text{CLIP}}(\text{image})$ (Figure \ref{fig:ablation} \textcolor{atm}{[14]}).

\subsection{Additional Details on Evaluation Metrics}
\label{app:metrics}
We use the following image similarity metrics:
\begin{itemize}
\item PixCorr is the pixel-level correlation between the ground-truth images and reconstructed images.
\item SSIM is the structural similarity index metric \cite{wang_image_2004}.
\item AlexNet($2$) and AlexNet($5$) are the 2-way comparisons (2WC) of layers 2 and 5 of AlexNet \cite{alexnet}.
\item CLIP is the 2WC of the output layer of the CLIP ViT-L/14 Vision model \cite{radford_learning_2021}.
\item Incep is the 2WC of the last pooling layer of InceptionV3 \cite{inceptionv3}.
\item Eff and SwAV are distance metrics gathered from EfficientNet-B13 \cite{tan_efficientnet_2020} and SwAV-ResNet50 \cite{caron_unsupervised_2021} models. 
\end{itemize}
For the metrics in Table $1$, a two-way comparison (2WC) evaluates whether the feature embedding of the stimulus image is more similar to the feature embedding of the target reconstruction, or the feature embedding of a randomly selected "distractor" reconstruction, where the score is the percent of correctly identified target reconstructions across a pool of "distractors". Our 2WC metrics, calculated using reconstructions of the $199$ other test-set stimuli as "distractors", have a notably different chance threshold from 2WC metrics presented in reconstruction papers that perform evaluations using a test set with a different number of "distractors", such as the shared1000 test set of NSD \cite{allen_massive_2022}, and are thus not directly comparable. All metrics in Table 1 were calculated and averaged across 10 images sampled from the output distribution of each method using a random seed. All metrics in Table were calculated on our reproduction of other methods using their open source code, and might differ slightly from metrics reported in the original papers due to our implementation of the metrics we calculated.

\subsection{Statistical Significance of Evaluation Metrics}
\label{app:statisticalsig}
\begin{table*}[!htb]
    \vspace{-10pt}
    \centering
    \setlength{\tabcolsep}{2pt}
    \small
    \resizebox{\textwidth}{!}{
    \begin{tabular}{lccccccccc}
        \toprule
        Method & \multicolumn{4}{c}{Low-Level} & \multicolumn{4}{c}{High-Level} & \multicolumn{1}{c}{Human Raters} \\
        \cmidrule(lr){2-5}\cmidrule(lr){6-9}\cmidrule(l){10-10}
        & PixCorr $\uparrow$ & SSIM $\uparrow$ & Alex(2) $\uparrow$ & Alex(5) $\uparrow$ & Incep $\uparrow$ & CLIP $\uparrow$ & Eff $\downarrow$ & SwAV $\downarrow$ & Ident.\ Acc.\ $\uparrow$ \\
        \midrule
        \multicolumn{10}{c}{\textbf{THINGS-EEG2}} \\
        \midrule
        ENIGMA (Multi-Subject) & $\pm$0.0014 & $\pm$0.0014 & $\pm$0.15\% & $\pm$0.12\% & $\pm$0.20\% & $\pm$0.19\% & $\pm$0.0008 & $\pm$0.0008 & $\pm$0.89\% \\ 
        ATM-S (Multi-Subject) & $\pm$0.0009 & $\pm$0.0010 & $\pm$0.20\% & $\pm$0.20\% & $\pm$0.21\% & $\pm$0.21\% & $\pm$0.0004 & $\pm$0.0006 & $\pm$1.15\% \\ 
        \midrule
        ENIGMA (Single-Subject) & $\pm$0.0014 & $\pm$0.0014 & $\pm$0.14\% & $\pm$0.11\% & $\pm$0.20\% & $\pm$0.18\% & $\pm$0.0008 & $\pm$0.0008 & $\pm$0.87\% \\ 
        ATM-S (Single-Subject) & $\pm$0.0013 & $\pm$0.0013 & $\pm$0.17\% & $\pm$0.15\% & $\pm$0.21\% & $\pm$0.20\% & $\pm$0.0007 & $\pm$0.0008 & $\pm$0.97\% \\ 
        Perceptogram (Single-Subject) & $\pm$0.0014 & $\pm$0.0015 & $\pm$0.12\% & $\pm$0.11\% & $\pm$0.20\% & $\pm$0.20\% & $\pm$0.0007 & $\pm$0.0007 & $\pm$0.94\% \\ 
        \midrule
        \multicolumn{10}{c}{\textbf{Alljoined-1.6M}} \\
        \midrule
        ENIGMA (Multi-Subject) & $\pm$0.0007 & $\pm$0.0008 & $\pm$0.14\% & $\pm$0.13\% & $\pm$0.15\% & $\pm$0.15\% & $\pm$0.0005 & $\pm$0.0005 & $\pm$0.77\% \\ 
        ATM-S (Multi-Subject) & $\pm$0.0007 & $\pm$0.0007 & $\pm$0.14\% & $\pm$0.15\% & $\pm$0.15\% & $\pm$0.15\% & $\pm$0.0003 & $\pm$0.0004 & $\pm$0.82\% \\ 
        \midrule
        ENIGMA (Single-Subject) & $\pm$0.0007 & $\pm$0.0009 & $\pm$0.14\% & $\pm$0.14\% & $\pm$0.15\% & $\pm$0.15\% & $\pm$0.0004 & $\pm$0.0005 & $\pm$0.78\% \\ 
        ATM-S (Single-Subject) & $\pm$0.0007 & $\pm$0.0008 & $\pm$0.14\% & $\pm$0.15\% & $\pm$0.15\% & $\pm$0.15\% & $\pm$0.0004 & $\pm$0.0005 & $\pm$0.80\% \\ 
        Perceptogram (Single-Subject) & $\pm$0.0008 & $\pm$0.0010 & $\pm$0.13\% & $\pm$0.13\% & $\pm$0.15\% & $\pm$0.15\% & $\pm$0.0004 & $\pm$0.0005 & $\pm$0.79\% \\ 
        \bottomrule
    \end{tabular}

    }
    \caption{Standard error measurements for evaluation metrics of EEG-to-Image reconstruction models evaluated on the THINGS‑EEG2 and Alljoined-1.6M datasets. Values correspond to the standard error spread of values in Table \ref{table:featuremetrics} in the manuscript.} 
    \label{table:statisticalsignificance}
\end{table*}
\FloatBarrier

\subsection{Median and Worst-case Reconstructions}
\label{app:median_worst}
\begin{figure}[!htb]
  \centering
  \includegraphics[width=\columnwidth]{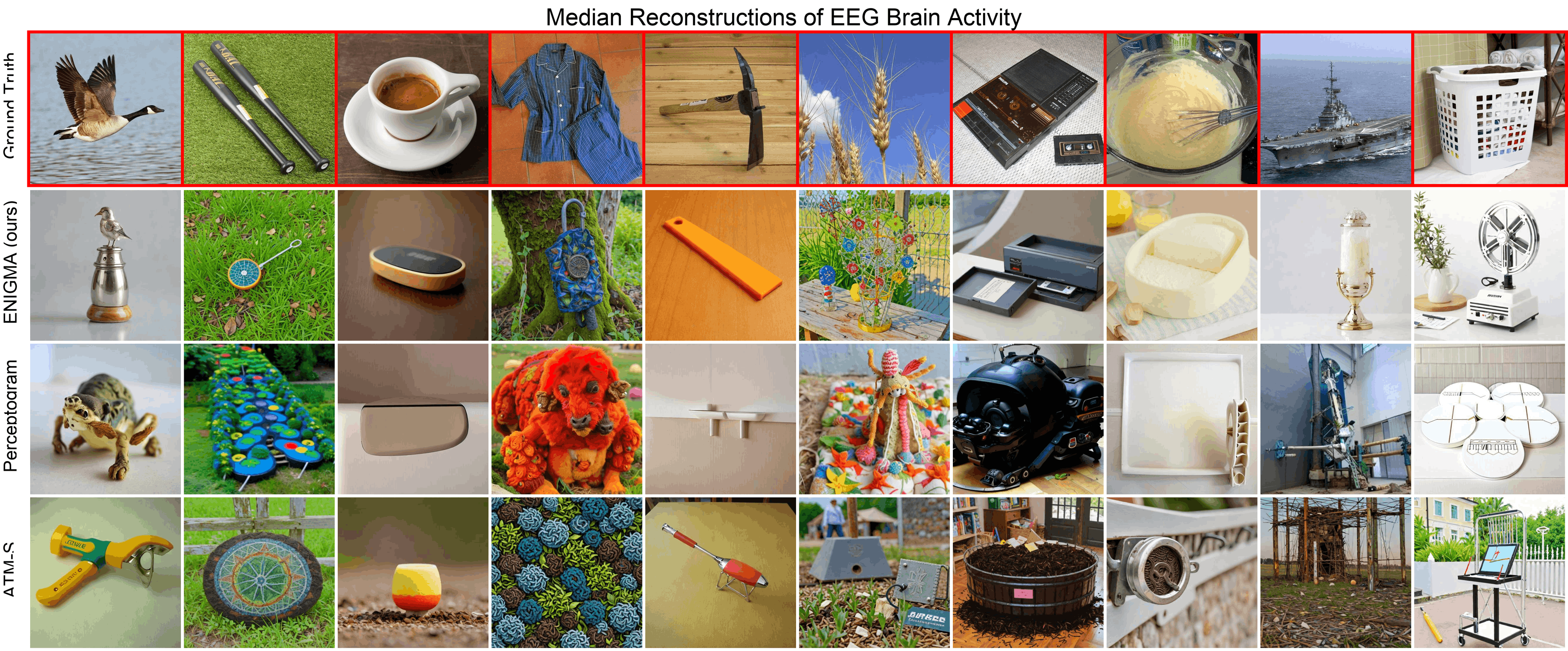}
  \caption{%
    Qualitative comparison of reconstruction methods on seen stimuli from THINGS-EEG2. Reconstructions selected are the outputs sampled from each method and stimulus with the median scores on all of the image feature metrics in Table \ref{table:featuremetrics}.}
  \label{fig:median_things}
\end{figure}

\begin{figure}[!htb]
  \centering
  \includegraphics[width=\columnwidth]{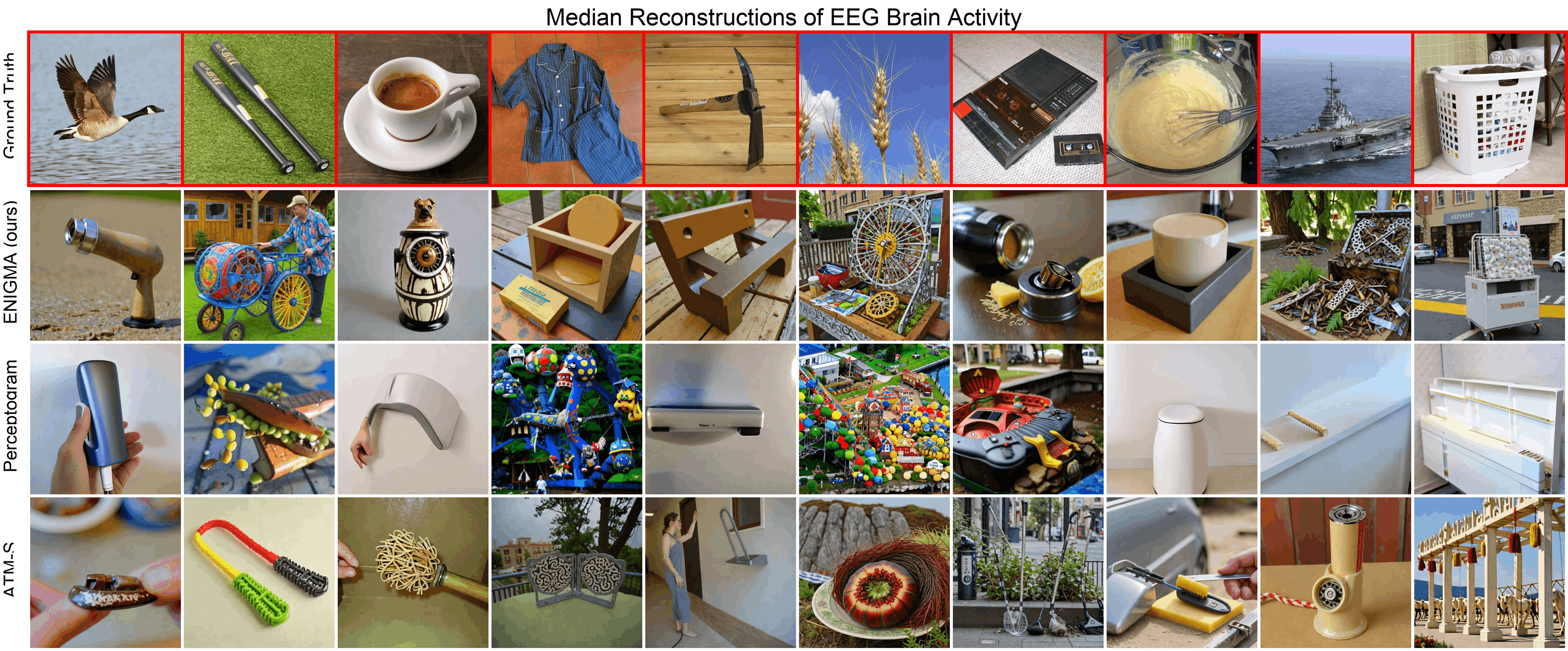}
  \caption{%
    Qualitative comparison of reconstruction methods on seen stimuli from Alljoined-1.6M. Reconstructions selected are the outputs sampled from each method and stimulus with the median scores on all of the image feature metrics in Table \ref{table:featuremetrics}.}
  \label{fig:median_aj}
\end{figure}

\begin{figure}[!htb]
  \centering
  \includegraphics[width=\columnwidth]{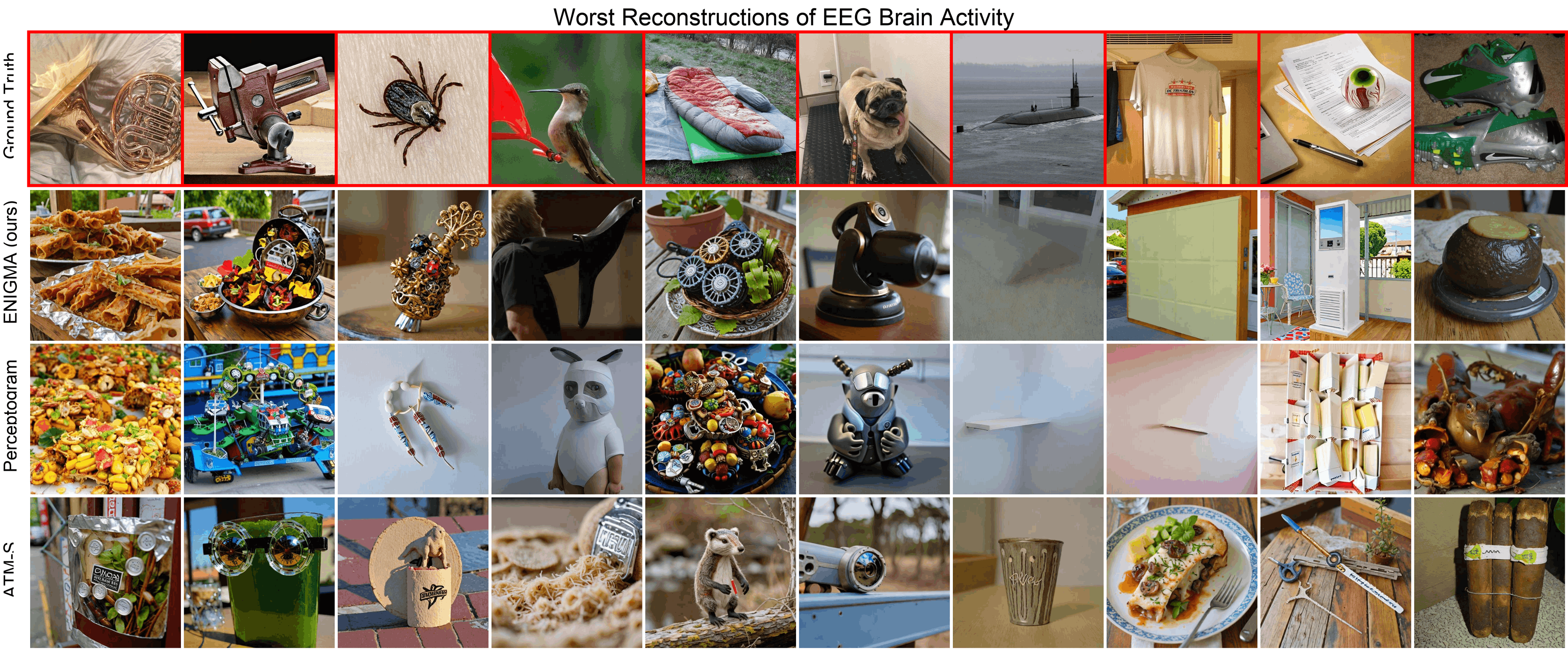}
  \caption{%
    Qualitative comparison of reconstruction methods on seen stimuli from THINGS-EEG2. Reconstructions selected are the outputs sampled from each method and stimulus with the worst scores on all of the image feature metrics in Table \ref{table:featuremetrics}.}
  \label{fig:worst_things}
\end{figure}

\begin{figure}[!htb]
  \centering
  \includegraphics[width=\columnwidth]{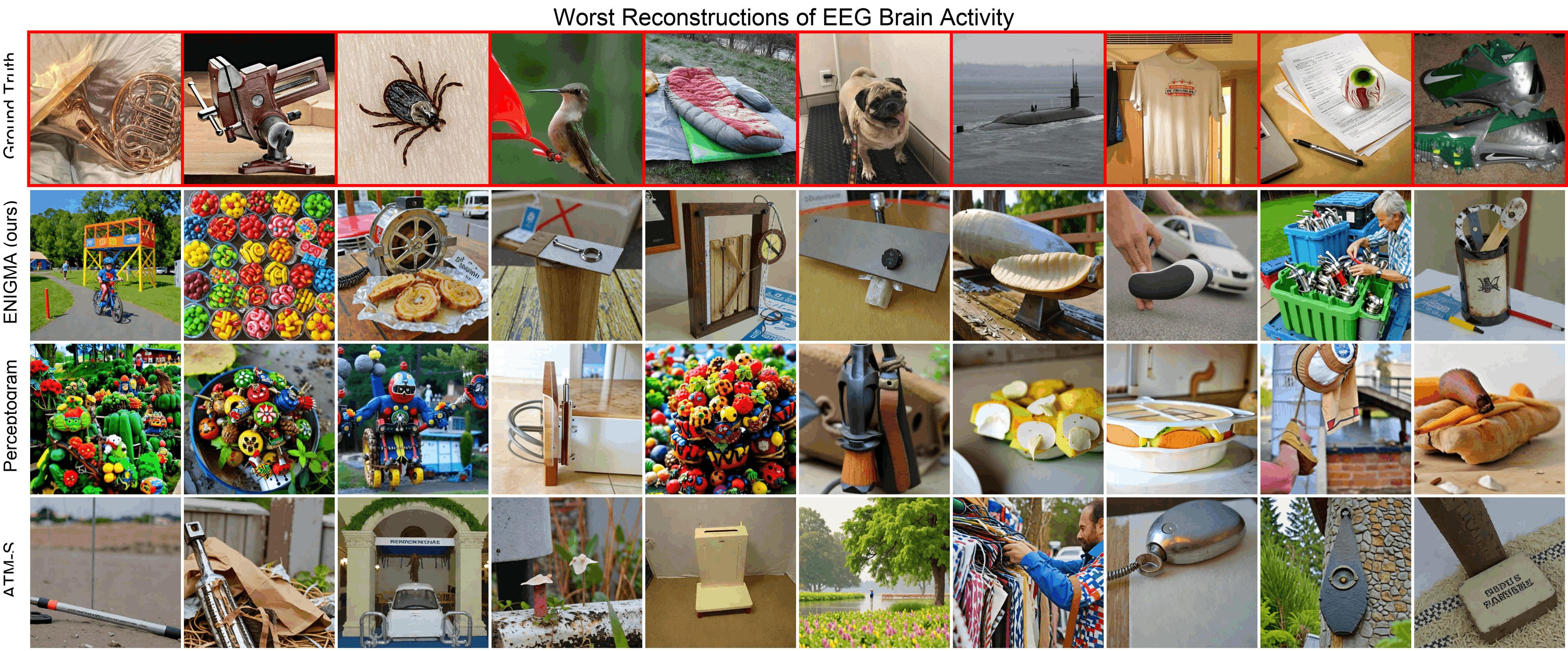}
  \caption{%
    Qualitative comparison of reconstruction methods on seen stimuli from Alljoined-1.6M. Reconstructions selected are the outputs sampled from each method and stimulus with the worst scores on all of the image feature metrics in Table \ref{table:featuremetrics}.}
  \label{fig:worst_aj}
\end{figure}
\FloatBarrier

\subsection{Dataset Scaling Analysis}
\label{app:scaling}

\begin{figure}[!htb]
\vspace{-5pt}
  \centering
  \includegraphics[width=0.5\columnwidth]{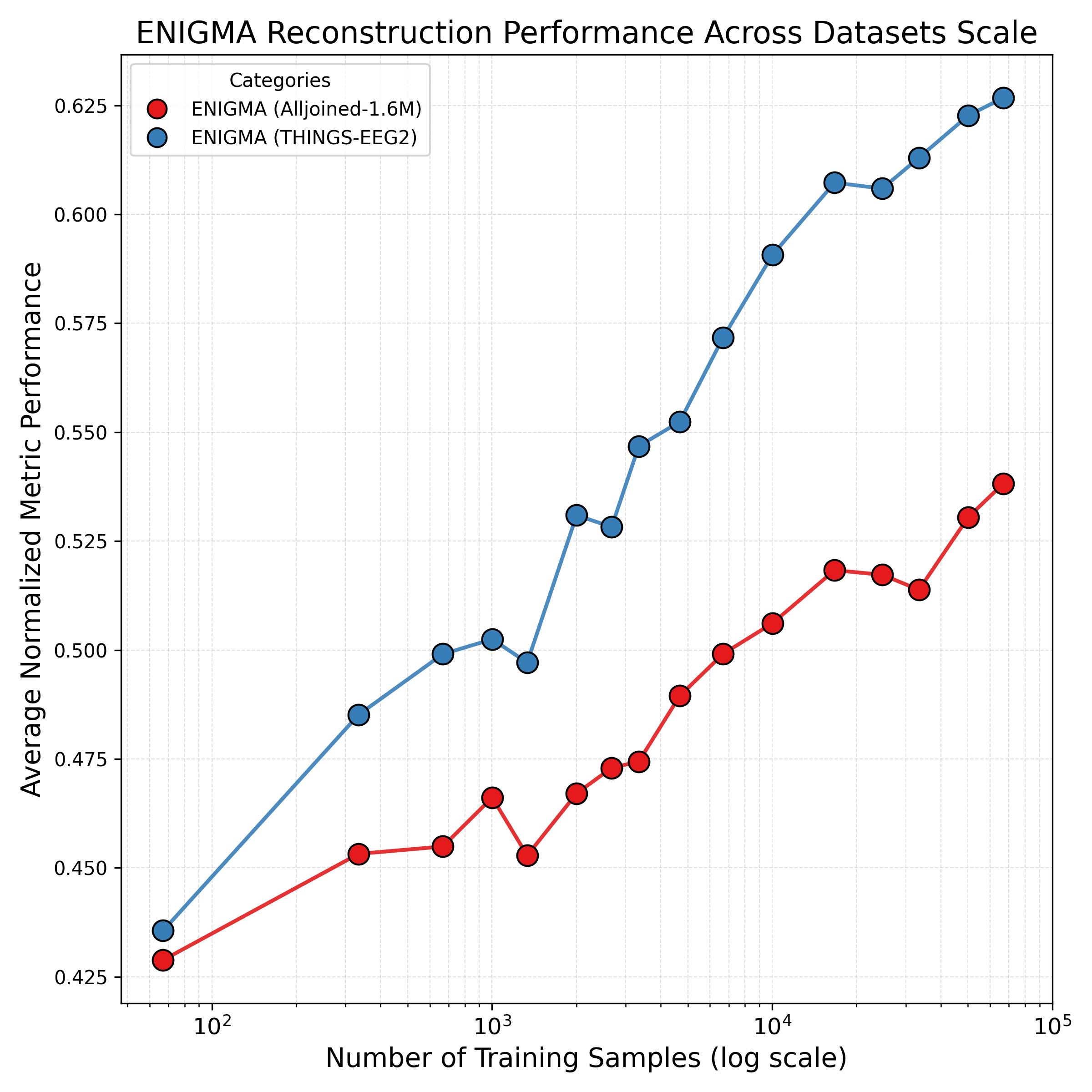}
  \vspace{-6pt}
  \caption{%
    Scaling analysis of \textbf{ENIGMA} (not pretrained) performance on the THINGS-EEG2 and Alljoined-1.6M datasets. The number of training samples are plotted on a log-scale X-axis, and the normalized average of feature metrics presented in Table \ref{table:featuremetrics} is plotted on the Y-axis.}
  \vspace{-8pt}
  \label{fig:scaling}
\end{figure}

When collecting large neuroimaging datasets, it is useful to be able to predict how efficiently models will scale on newly collected data. As shown in Figure \ref{fig:scaling}, the reconstruction performance of \textbf{ENIGMA} increases log-linearly with the number of training samples on both available datasets, with no evident saturation on either dataset, however, performance scales with a larger exponent factor on THINGS‑EEG2 that was collected on research-grade hardware. Such divergence suggests that while sheer data volume does reliably boosts accuracy, the quality of recording hardware significantly accelerates learning efficiency and leaves headroom for further gains. As highlighted with the release of Alljoined-1.6M \cite{Alljoined-1.6M}, this difference in scaling efficiency between EEG hardware quality is a key limitation to overcome for practical BCI applications.
\FloatBarrier

\subsection{Channel Count Ablation Analysis}
\label{app:channelcount}

\begin{figure}[!htb]
\centering
\includegraphics[width=0.5\textwidth]{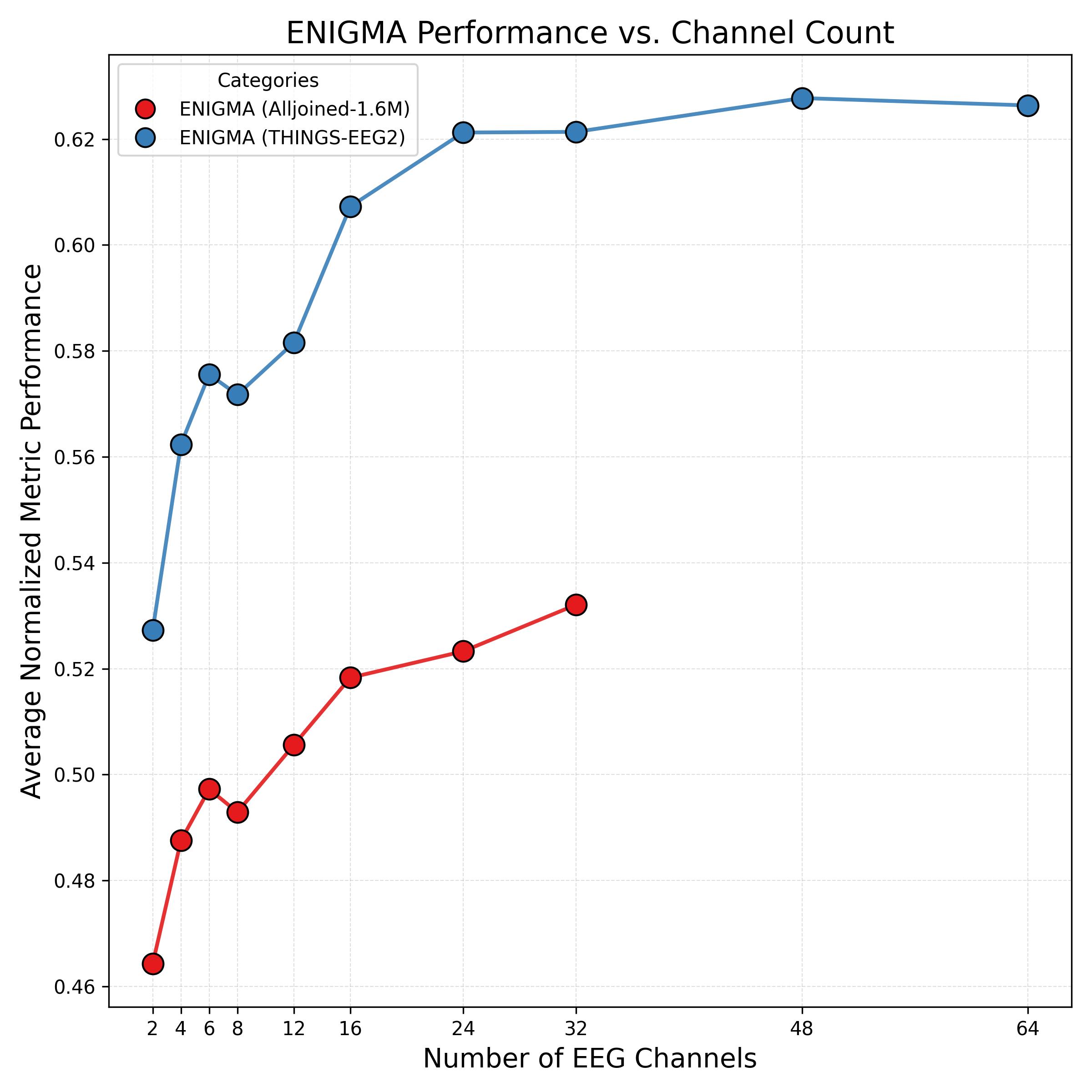}
\vspace{-5pt}
\caption{Channel count analysis of model performance for each dataset. The number of channels in each dataset was progressively reduced, while the remaining channels focus primarily on occipital cortex. The Y axis is plotted the same as Fig. \ref{fig:ablation}B.}
\label{figure:channelscaling}
\end{figure}

A commonly-asked question is the number of channels needed to obtain high quality reconstructions from EEG-to-Image reconstruction models. We analyzed how the number of channels affects decoding performance using \textbf{ENIGMA}, and explored whether this contributed significantly to differences in performance between performance on the two benchmark datasets. We sub-sampled varying numbers of channels from both datasets, while retaining a focus on covering occipital cortex. We find that while performance did drop with fewer channels, channel count is not the most significant factor accounting for the performance difference between the datasets, and performance starts to drop off after 24 channels for both datasets. This suggest that it might be possible to achieve reasonable decoding performance with fewer than 32 channels.
\FloatBarrier
\subsection{Behavioral Evaluation Experiments}
\label{app:behavioral}
To evaluate the quality of EEG-to-Image reconstruction models applied to our dataset, we conducted a behavioral experiment on $545$ human raters online. For our experiment, we identified no risks to the human participants, and collected informed consent from all participants. 

The experiment stimuli consists of image reconstruction sampled from the 30 subjects across THINGS-EEG2 and Alljoined-1.6M from all methods and cases in Table \ref{table:featuremetrics}. The images were shuffled and $60$ images presented to each subject. We use attention checks to identify whether human raters were paying attention to the task and the instructions and dropped $8$ human raters who failed at least $2$ out $8$  attentions checks before analysis. An attention check presents the ground truth image as one of the candidate images and raters have to select the candidate ground truth image (as an image is most similar to itself) to pass.

Our subjects were recruited through the \href{http://www.prolific.ac.uk)}{Prolific platform}, with our experimental tasks hosted on \href{http://meadows-research.com}{Meadows}. Each human rater was paid $\$1.25$ for the completion of the experiment, and the median completion time was $5$ minutes, resulting in an average payment rate of $\$15$/hour. The code to reproduce our experiment can be found in \href{https://anonymous.4open.science/r/ENIGMA-9C46/}{our anonymized GitHub repository.}

\subsubsection{2AFC Identification Task}
\begin{figure}[!htb]
\begin{center}
\includegraphics[width=\columnwidth]{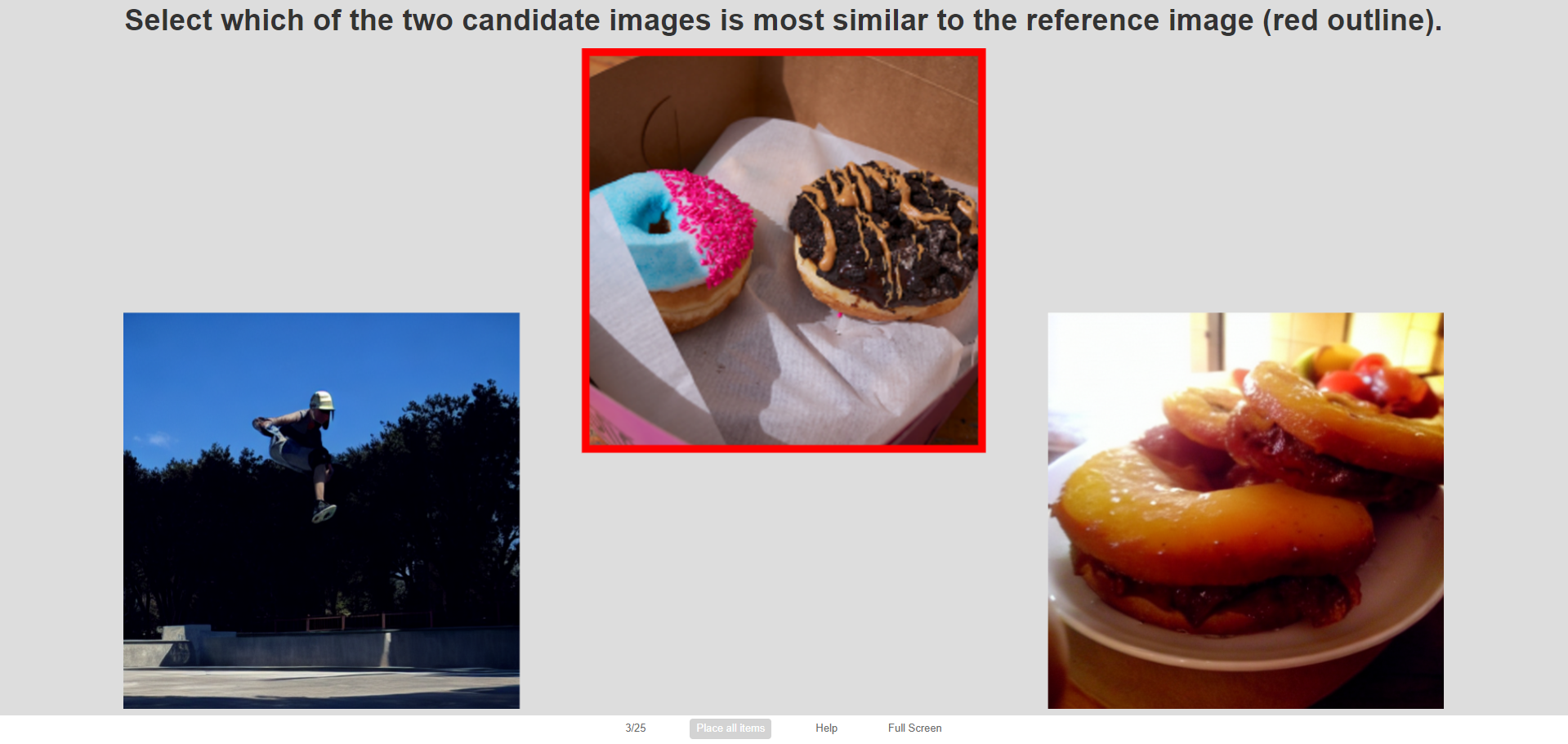}
\end{center}
\caption{An example of the $2$ alternative forced choice task used in our behavioral experiment performed by human raters.} 
\label{figure:task1}
\end{figure}

Our experiment was a $2$ alternative forced choice task (2AFC) facilitated by the "Match-To-Sample" task on the Meadows platform. An example of the first experiment can be seen in Figure \ref{figure:task1}. In this experiment, human raters were asked to select which of two candidate images was more similar to a reference image. The reference image provided is the ground truth image the subject either saw, and the $2$ candidate images were the target reconstruction of the reference image, or a randomly selected reconstruction from an EEG recording corresponding to a different stimulus. The two candidate images were always sampled from the same reconstruction method and subject. This experiment was repeated for all reconstruction methods, model types, datasets, and subjects. With the results presented in Table \ref{table:featuremetrics}, we establish a baseline for human-rated image identification accuracy of seen image reconstructions from EEG, as no other paper has conducted behavioral evaluations of EEG-to-Image reconstructions.

\bibliographystyle{icml2026}
\bibliography{main}

@String(CVPR= {IEEE Conf. Comput. Vis. Pattern Recog.})

@String(ICLR = {Int. Conf. Learn. Represent.})

@String(CVPR  = {CVPR})

@String(ICLR  = {ICLR})

@article{Gramfort2013,
  title={MEG and EEG data analysis with MNE-Python},
  author={Gramfort, Alexandre and Luessi, Martin and Larson, Eric and Engemann, Denis A. and Strohmeier, Daniel and Brodbeck, Christian and et al.},
  journal={Frontiers in Neuroscience},
  volume={7},
  pages={267},
  year={2013},
  doi={10.3389/fnins.2013.00267}
}

@article{badcock2013validation,
  title={Validation of the Emotiv EPOC{\textregistered} EEG gaming system for measuring research quality auditory ERPs},
  author={Badcock, Nicholas A and Mousikou, Petroula and Mahajan, Yatin and De Lissa, Peter and Thie, Johnson and McArthur, Genevieve},
  journal={PeerJ},
  volume={1},
  pages={e38},
  year={2013},
  publisher={PeerJ Inc.}
}

@article{williams2020validation,
  title={A validation of Emotiv EPOC Flex saline for EEG and ERP research},
  author={Williams, Nikolas S and McArthur, Genevieve M and de Wit, Bianca and Ibrahim, George and Badcock, Nicholas A},
  journal={PeerJ},
  volume={8},
  pages={e9713},
  year={2020},
  publisher={PeerJ Inc.}
}

@article{Guggenmos2018,
  title={Multivariate pattern analysis for MEG: A comparison of dissimilarity measures},
  author={Guggenmos, Matthias and Sterzer, Philipp and Cichy, Radoslaw M},
  journal={NeuroImage},
  volume={173},
  pages={434--447},
  year={2018},
  publisher={Elsevier}
}

@inproceedings{takagi2022_decoding,
  title={High-resolution image reconstruction with latent diffusion models from human brain activity},
  author={Takagi, Yu and Nishimoto, Shinji},
  booktitle={Proceedings of the IEEE/CVF Conference on Computer Vision and Pattern Recognition},
  pages={14453--14463},
  year={2023}
}

@misc{takagi2023improving,
      title={Improving visual image reconstruction from human brain activity using latent diffusion models via multiple decoded inputs}, 
      author={Takagi, Yu and Nishimoto, Shinji},
      year={2023},
      eprint={2306.11536},
      archivePrefix={arXiv},
      primaryClass={q-bio.NC}
}

@article{ozcelik2023braindiffuser,
  title={Natural scene reconstruction from {fMRI} signals using generative latent diffusion},
  author={Furkan Ozcelik and Rufin VanRullen},
  journal={Scientific Reports},
  year={2023},
  volume={13},
  url={https://api.semanticscholar.org/CorpusID:260439960}
}

@article{BREEDLOVE20202211,
title = {Generative Feedback Explains Distinct Brain Activity Codes for Seen and Mental Images},
journal = {Current Biology},
volume = {30},
number = {12},
pages = {2211-2224.e6},
year = {2020},
issn = {0960-9822},
doi = {https://doi.org/10.1016/j.cub.2020.04.014},
url = {https://www.sciencedirect.com/science/article/pii/S0960982220304942},
author = {Jesse L. Breedlove and Ghislain St-Yves and Cheryl A. Olman and Thomas Naselaris}
}

@inproceedings{kneeland2023reconstructing,
      title={Reconstructing seen images from human brain activity via guided stochastic search}, 
      author={Reese Kneeland and Jordyn Ojeda and Ghislain St-Yves and Thomas Naselaris},
      doi = {10.32470/CCN.2023.1672-0},
      url = {https://2023.ccneuro.org/view_paper1337.html?PaperNum=1672},
      booktitle = {Conference on Cognitive Computational Neuroscience},
      year={2023},
      eprint={2305.00556},
      archivePrefix={arXiv},
      primaryClass={q-bio.NC}
}

@misc{kneeland_second_2023,
	title = {Second {Sight}: {Using} brain-optimized encoding models to align image distributions with human brain activity},
	shorttitle = {Second {Sight}},
	url = {http://arxiv.org/abs/2306.00927},
	doi = {10.48550/arXiv.2306.00927},
	urldate = {2024-01-16},
	publisher = {arXiv},
	author = {Kneeland, Reese and Ojeda, Jordyn and St-Yves, Ghislain and Naselaris, Thomas},
	month = jun,
	year = {2023},
	note = {arXiv:2306.00927 [cs, q-bio]},
}

@misc{kneeland_brain-optimized_2023,
	title = {Brain-optimized inference improves reconstructions of {fMRI} brain activity},
	url = {http://arxiv.org/abs/2312.07705},
	doi = {10.48550/arXiv.2312.07705},
	urldate = {2024-01-13},
	publisher = {arXiv},
	author = {Kneeland, Reese and Ojeda, Jordyn and St-Yves, Ghislain and Naselaris, Thomas},
	month = dec,
	year = {2023},
	note = {arXiv:2312.07705 [cs, q-bio]},
}

@inproceedings{sauer2024adversarial,
  title={Adversarial diffusion distillation},
  author={Sauer, Axel and Lorenz, Dominik and Blattmann, Andreas and Rombach, Robin},
  booktitle={European Conference on Computer Vision},
  pages={87--103},
  year={2024},
  organization={Springer}
}

@inproceedings{
scotti_reconstructing_2023,
title={Reconstructing the Mind's Eye: f{MRI}-to-Image with Contrastive Learning and Diffusion Priors},
author={Paul Steven Scotti and Atmadeep Banerjee and Jimmie Goode and Stepan Shabalin and Alex Nguyen and Cohen Ethan and Aidan James Dempster and Nathalie Verlinde and Elad Yundler and David Weisberg and Kenneth Norman and Tanishq Mathew Abraham},
booktitle={Thirty-seventh Conference on Neural Information Processing Systems},
year={2023},
url={https://openreview.net/forum?id=rwrblCYb2A}
}

@inproceedings{tan_efficientnet_2020,
  author       = {Mingxing Tan and
                  Quoc V. Le},
  editor       = {Kamalika Chaudhuri and
                  Ruslan Salakhutdinov},
  title        = {EfficientNet: Rethinking Model Scaling for Convolutional Neural Networks},
  booktitle    = {Proceedings of the 36th International Conference on Machine Learning,
                  {ICML} 2019, 9-15 June 2019, Long Beach, California, {USA}},
  series       = {Proceedings of Machine Learning Research},
  volume       = {97},
  pages        = {6105--6114},
  publisher    = {{PMLR}},
  year         = {2019},
  url          = {http://proceedings.mlr.press/v97/tan19a.html},
  bibsource    = {dblp computer science bibliography, https://dblp.org}
}

@article{caron_unsupervised_2021,
  author       = {Mathilde Caron and
                  Ishan Misra and
                  Julien Mairal and
                  Priya Goyal and
                  Piotr Bojanowski and
                  Armand Joulin},
  title        = {Unsupervised Learning of Visual Features by Contrasting Cluster Assignments},
  journal      = {CoRR},
  volume       = {abs/2006.09882},
  year         = {2020},
  url          = {https://arxiv.org/abs/2006.09882},
  eprinttype    = {arXiv},
  eprint       = {2006.09882},
  bibsource    = {dblp computer science bibliography, https://dblp.org}
}

@article{wang_image_2004,
	title = {Image quality assessment: from error visibility to structural similarity},
	volume = {13},
	issn = {1941-0042},
	shorttitle = {Image quality assessment},
	doi = {10.1109/TIP.2003.819861},
	number = {4},
	journal = {IEEE Transactions on Image Processing},
	author = {Wang, Zhou and Bovik, A.C. and Sheikh, H.R. and Simoncelli, E.P.},
	month = apr,
	year = {2004},
	note = {Conference Name: IEEE Transactions on Image Processing},
	pages = {600--612},
}

@article{srivastava2014dropout,
  title={Dropout: a simple way to prevent neural networks from overfitting},
  author={Srivastava, Nitish and Hinton, Geoffrey and Krizhevsky, Alex and Sutskever, Ilya and Salakhutdinov, Ruslan},
  journal={The journal of machine learning research},
  volume={15},
  number={1},
  pages={1929--1958},
  year={2014},
  publisher={JMLR. org}
}

@misc{radford_learning_2021,
	title = {Learning {Transferable} {Visual} {Models} {From} {Natural} {Language} {Supervision}},
	url = {http://arxiv.org/abs/2103.00020},
	doi = {10.48550/arXiv.2103.00020},
	urldate = {2023-04-30},
	publisher = {arXiv},
	author = {Radford, Alec and Kim, Jong Wook and Hallacy, Chris and Ramesh, Aditya and Goh, Gabriel and Agarwal, Sandhini and Sastry, Girish and Askell, Amanda and Mishkin, Pamela and Clark, Jack and Krueger, Gretchen and Sutskever, Ilya},
	month = feb,
	year = {2021},
	note = {arXiv:2103.00020 [cs]},
}

@article{allen_massive_2022,
	title = {A massive {7T} {fMRI} dataset to bridge cognitive neuroscience and artificial intelligence},
	volume = {25},
	issn = {1097-6256, 1546-1726},
	url = {https://www.nature.com/articles/s41593-021-00962-x},
	doi = {10.1038/s41593-021-00962-x},
	language = {en},
	number = {1},
	urldate = {2022-11-01},
	journal = {Nature Neuroscience},
	author = {Allen, Emily J. and St-Yves, Ghislain and Wu, Yihan and Breedlove, Jesse L. and Prince, Jacob S. and Dowdle, Logan T. and Nau, Matthias and Caron, Brad and Pestilli, Franco and Charest, Ian and Hutchinson, J. Benjamin and Naselaris, Thomas and Kay, Kendrick},
	month = jan,
	year = {2022},
	pages = {116--126},
}

@inproceedings{alexnet,
 author = {Krizhevsky, Alex and Sutskever, Ilya and Hinton, Geoffrey E},
 booktitle = {Advances in Neural Information Processing Systems},
 editor = {F. Pereira and C.J. Burges and L. Bottou and K.Q. Weinberger},
 pages = {},
 publisher = {Curran Associates, Inc.},
 title = {ImageNet Classification with Deep Convolutional Neural Networks},
 url = {https://proceedings.neurips.cc/paper_files/paper/2012/file/c399862d3b9d6b76c8436e924a68c45b-Paper.pdf},
 volume = {25},
 year = {2012}
}

@article{inceptionv3,
  author       = {Christian Szegedy and
                  Vincent Vanhoucke and
                  Sergey Ioffe and
                  Jonathon Shlens and
                  Zbigniew Wojna},
  title        = {Rethinking the Inception Architecture for Computer Vision},
  journal      = {CoRR},
  volume       = {abs/1512.00567},
  year         = {2015},
  url          = {http://arxiv.org/abs/1512.00567},
  eprinttype    = {arXiv},
  eprint       = {1512.00567},
  bibsource    = {dblp computer science bibliography, https://dblp.org}
}

@article{peceptualsimilarity,
author = {Pawan Sinha and Richard Russell},
title ={A Perceptually Based Comparison of Image Similarity Metrics},

journal = {Perception},
volume = {40},
number = {11},
pages = {1269-1281},
year = {2011},
doi = {10.1068/p7063},
    note ={PMID: 22416586},

URL = { 
    
        https://doi.org/10.1068/p7063
    
    

},
eprint = { 
    
        https://doi.org/10.1068/p7063
    
    

}
}

@inproceedings{pickapic,
title={Pick-a-Pic: An Open Dataset of User Preferences for Text-to-Image Generation},
author={Yuval Kirstain and Adam Polyak and Uriel Singer and Shahbuland Matiana and Joe Penna and Omer Levy},
booktitle={Thirty-seventh Conference on Neural Information Processing Systems},
year={2023},
url={https://openreview.net/forum?id=G5RwHpBUv0}
}

@inproceedings{Scotti2024MindEye2,
author = {Scotti, Paul S. and Tripathy, Mihir and Villanueva, Cesare Kadir Torrico and Kneeland, Reese and Chen, Tong and Narang, Ashutosh and Santhirasegaran, Charan and Xu, Jonathan and Naselaris, Thomas and Norman, Kenneth A. and Abraham, Tanishq Mathew},
title = {MindEye2: shared-subject models enable fMRI-to-image with 1 hour of data},
year = {2024},
booktitle = {Proceedings of the 41st International Conference on Machine Learning},
articleno = {1794},
numpages = {22},
location = {Vienna, Austria},
}

@article{pearson2015mental,
  title={Mental imagery: functional mechanisms and clinical applications},
  author={Pearson, Joel and Naselaris, Thomas and Holmes, Emily A and Kosslyn, Stephen M},
  journal={Trends in cognitive sciences},
  volume={19},
  number={10},
  pages={590--602},
  year={2015},
  publisher={Elsevier}
}

@article{holmes2010mental,
  title={Mental imagery in emotion and emotional disorders},
  author={Holmes, Emily A and Mathews, Andrew},
  journal={Clinical psychology review},
  volume={30},
  number={3},
  pages={349--362},
  year={2010},
  publisher={Elsevier}
}

@inproceedings{NSDImagery,
    title = {{NSD-Imagery}: A benchmark dataset for extending {fMRI} vision decoding methods to mental imagery},
    author = {Reese Kneeland and Paul S. Scotti and Ghislain St-Yves and Jesse Breedlove and Kendrick Kay and Thomas Naselaris},
    booktitle = {Proceedings of the IEEE/CVF Conference on Computer Vision and Pattern Recognition (CVPR)},
    month     = {June},
    year      = {2025},
}

@inproceedings{Spampinato2017,
  title={{Deep Learning Human Mind for Automated Visual Classification}},
  author={Spampinato, Carlo and Palazzo, Sebastiano and Kavasidis, Ignazio and Giordano, Daniele and Souly, Nada and Shah, Mubarak},
  booktitle={Proceedings of the IEEE Conference on Computer Vision and Pattern Recognition (CVPR)},
  year={2017},
  pages={6809--6818}
}

@article{Li2020,
  title={{Training on the test set? An analysis of Spampinato et al.'s EEG image classification method}},
  author={Li, Ren and Johansen, Jared S. and Ahmed, Hamad and Ilyevsky, Thomas V. and Wilbur, Ronnie B. and Bharadwaj, Hari M. and Siskind, Jeffrey M.},
  journal={IEEE Trans. Pattern Analysis and Machine Intelligence},
  year={2020},
  note={\textit{(Early Access) arXiv:1812.07697}}
}

@article{Grootswagers2022,
  title={{Human EEG recordings for 1,854 concepts presented in rapid serial visual presentation streams}},
  author={Grootswagers, Tijl and Zhou, Ivy and Robinson, Amanda K. and Hebart, Martin N. and Carlson, Thomas A.},
  journal={Scientific Data},
  volume={9},
  number={3},
  year={2022},
  doi={10.1038/s41597-021-01102-7}
}

@article{Gifford2022,
  title={{A large and rich EEG dataset for modeling human visual object recognition}},
  author={Gifford, Alessandro T. and Dwivedi, Kshitij and Roig, Gemma and Cichy, Radoslaw M.},
  journal={NeuroImage},
  volume={264},
  pages={119754},
  year={2022},
  doi={10.1016/j.neuroimage.2022.119754}
}

@inproceedings{Song2024,
  title={{Decoding Natural Images from EEG for Object Recognition}},
  author={Song, Yonghao and Liu, Bingchuan and Li, Xiang and Shi, Nanlin and Wang, Yijun and Gao, Xiaorong},
  booktitle={Proceedings of the International Conference on Learning Representations (ICLR)},
  year={2024}
}

@article{Fei2024-perceptogram,
  title={{Perceptogram: Reconstructing Visual Percepts from EEG}},
  author={Fei, Teng and Uppal, Abhinav and Jackson, Ian and Ravishankar, Srinivas and Wang, David and de Sa, Virginia R.},
  journal={arXiv preprint arXiv:2404.01250},
  year={2024},
  note={(extended version with additional analyses)}
}

@inproceedings{Li2024,
  title={{Visual Decoding and Reconstruction via EEG Embeddings with Guided Diffusion}},
  author={Li, Dongyang and Wei, Chen and Li, Shiying and Zou, Jiachen and Liu, Quanying},
  booktitle={Advances in Neural Information Processing Systems (NeurIPS)},
  year={2024}
}

@article{Ye2023,
  title={{IP-Adapter: Text Compatible Image Prompt Adapter for Text-to-Image Diffusion Models}},
  author={Ye, Hu and Zhang, Jun and Liu, Sibo and Han, Xiao and Yang, Wei},
  journal={arXiv preprint arXiv:2308.06721},
  year={2023}
}

@misc{Alljoined-1.6M,
      title={Alljoined-1.6M: A Million-Trial EEG-Image Dataset for Evaluating Affordable Brain-Computer Interfaces}, 
      author={Jonathan Xu and Ugo Bruzadin Nunes and Wangshu Jiang and Samuel Ryther and Jordan Pringle and Paul S. Scotti and Arnaud Delorme and Reese Kneeland},
      year={2025},
      eprint={2508.18571},
      archivePrefix={arXiv},
      primaryClass={q-bio.NC},
      url={https://arxiv.org/abs/2508.18571}, 
}

@misc{InfoNCE,
      title={Representation Learning with Contrastive Predictive Coding}, 
      author={Aaron van den Oord and Yazhe Li and Oriol Vinyals},
      year={2019},
      eprint={1807.03748},
      archivePrefix={arXiv},
      primaryClass={cs.LG},
      url={https://arxiv.org/abs/1807.03748}, 
}

@article{sethics,
    doi = {10.1371/journal.pbio.3002899},
    author = {Gordon, Emma C. AND Seth, Anil K.},
    journal = {PLOS Biology},
    publisher = {Public Library of Science},
    title = {Ethical considerations for the use of brain–computer interfaces for cognitive enhancement},
    year = {2024},
    month = {10},
    volume = {22},
    url = {https://doi.org/10.1371/journal.pbio.3002899},
    pages = {1-15},
    number = {10},
}

@article{banville2025scaling,
  title={Scaling laws for decoding images from brain activity},
  author={Banville, Hubert and Benchetrit, Yohann and d'Ascoli, St{\'e}phane and Rapin, J{\'e}r{\'e}my and King, Jean-R{\'e}mi},
  journal={arXiv preprint arXiv:2501.15322},
  year={2025}
}

\end{document}